\documentclass[a4paper,11pt]{article}
\pdfoutput=1

\usepackage{jcappub} 
\usepackage[T1]{fontenc} 
\usepackage{braket}
\usepackage{xcolor}
\usepackage{tikz}
\usepackage{cancel}
\usepackage{comment}
\usepackage[normalem]{ulem}
\usepackage{cleveref}
\usepackage{bm}
\usepackage{arydshln}

\crefname{figure}{figure}{figures}
\Crefname{figure}{Figure}{Figures}

\newcommand{\vg}{``}

\newcommand{\dnef}{\Delta N_{\rm eff}}
\newcommand{\class}{\texttt{CLASS}}
\newcommand{\calN}{{\mathcal N}}
\newcommand{\pmin}{p_\star^{\rm min}}

\title{Cosmological Bounds on Free-Streaming Axion-like Particles from Cosmic Strings}

\author[a,b]{Riccardo Impavido,}
\author[c,b]{Nicola Barbieri,}
\author[d]{Luca Caloni,}
\author[b]{Massimiliano Lattanzi,}
\author[e,f]{and Luca Visinelli}

\affiliation[a]{Dipartimento di Fisica e Scienze della Terra, Universit\`a degli Studi di Ferrara, via Giuseppe Saragat 1, 44122 Ferrara, Italy}
\affiliation[b]{INFN, Sezione di Ferrara, via Giuseppe Saragat 1, 44122 Ferrara, Italy}
\affiliation[c]{Instituto de Física Corpuscular (IFIC), CSIC‐Universitat de València, Parc Científic UV, c/ Catedrático José Beltrán, 2, E-46980 Paterna (València), Spain}
\affiliation[d]{CFisUC, Department of Physics, University of Coimbra, P-3004-516, Coimbra, Portugal}
\affiliation[e]{Dipartimento di Fisica ``E.R. Caianiello'', Universit\`a degli Studi di Salerno, Via Giovanni Paolo II 132, 84084 Fisciano (SA), Italy}
\affiliation[f]{Istituto Nazionale di Fisica Nucleare, Gruppo Collegato di Salerno, Sezione di Napoli, Via Giovanni Paolo II 132, 84084 Fisciano (SA), Italy}

\emailAdd{mpvrcr@unife.it}
\emailAdd{barbieri@ific.uv.es}
\emailAdd{luca.caloni@uc.pt}
\emailAdd{lattanzi@fe.infn.it}
\emailAdd{lvisinelli@unisa.it}

\abstract{Axion-like particles (ALPs) radiated from a network of cosmic strings have been proposed as candidates that could account for all of the dark matter. They are characterized by a highly non-thermal, almost scale-invariant phase-space distribution, which spans many orders of magnitude in momentum. Building on results from existing numerical simulations of string networks, we implement ALPs from strings in the \texttt{CLASS} code to compute predictions for cosmological observables, such as the spectrum of cosmic microwave background (CMB) anisotropies. We exploit CMB data from {\it Planck}, the Atacama Cosmology Telescope and the South Pole Telescope, together with baryon acoustic oscillation measurements from the Dark Energy Spectroscopic Instrument, to constrain their abundance, for ALP masses $10^{-21} \, \mathrm{eV} \le m_a \le 10^{-12} \, \mathrm{eV}$. Constraints are driven by the effect of free-streaming ALPs on the lensing of CMB anisotropies. Using conservative data cuts that severely limit the potential impact of nonlinear corrections to theoretical predictions, we find that ALPs from cosmic strings cannot account for more than few per cent of dark matter for $m_a \le 10^{-20} \, \mathrm{eV}$, and cannot be the most abundant dark matter component for $m_a \le 10^{-18} \, \mathrm{eV}$.
}

\begin{document}

\maketitle

\section{Introduction}

    The nature of dark matter (DM) remains one of the major open questions in cosmology and particle physics. Among the proposed candidates, axions and axion-like particles (ALPs) are particularly well motivated. The QCD axion was originally introduced to solve the strong CP problem of the Standard Model~\cite{Peccei:1977hh, Weinberg:1977ma, Wilczek:1977pj}. More general ALPs can arise as pseudo Nambu-Goldstone bosons associated with spontaneously broken global symmetries~\cite{Srednicki:1985xd, Georgi:1986df, Kim:2015yna, Choi:2020rgn, Irastorza:2018dyq, Arza:2026rsl}, or emerge from the compactification of higher dimensional gauge fields in string theory~\cite{Svrcek:2006yi, Arvanitaki:2009fg}. Unlike the QCD axion, whose mass is related to the symmetry breaking scale~\cite{DiLuzio:2020wdo}, generic ALPs can have an independent mass $m_a$ and decay constant $f_a$, allowing for a broader range of cosmological scenarios~\cite{Arias:2012az,Marsh:2015xka}.

    The relic abundance and momentum distribution of axions and ALPs depend on their production mechanism and the expansion history of the early Universe~\cite{Grin:2007yg, Visinelli:2009kt}. Besides thermal production through interactions with the primordial plasma~\cite{Turner:1986tb, Masso:2002np, Graf:2010tv, Salvio:2013iaa, Ferreira:2018vjj, DEramo:2018vss, Arias-Aragon:2020shv, DEramo:2021psx, Caloni:2022uya, Badziak:2024szg, DEramo:2024jhn, Caloni:2024olo, Badziak:2025mkt, Barbieri:2026ewj}, axions and ALPs can be produced nonthermally through the misalignment mechanism~\cite{Abbott:1982af, Dine:1982ah, Preskill:1982cy} or through radiation from a network of cosmic strings~\cite{Davis:1986xc, Kibble:1976sj, Gorghetto:2018myk, Buschmann:2021juv, Saikawa:2024bta}. More generally, cosmic strings can also produce light gauge bosons, which provide additional cosmological signatures~\cite{Cheng:2024axj}. In scenarios where the underlying global symmetry breaks after inflation, the string network continuously emits ALPs until the formation of domain walls, eventually leading to its collapse for a domain wall number equal to one~\cite{Gorghetto:2018myk, Gorghetto:2021fsn}. The cosmological implications of axionic string networks have also been explored in connection with nonstandard expansion histories and gravitational wave signatures~\cite{Ramberg:2019dgi, Ramberg:2020oct, Chang:2021afa, Servant:2023mwt, Litos:2026npz}.

    Numerical simulations of cosmic string networks have extensively investigated ALP production and the evolution of the emitted spectra~\cite{Klaer:2017ond, Hindmarsh:2020pvl, Hindmarsh:2021vih, Gorghetto:2018myk, OHare:2021zrq, Gorghetto:2021fsn, Kim:2024dtq, Drew:2023ptp, Battye:2026whd, Kaltschmidt:2026hpj}. The instantaneous ALP emission spectrum approximately follows a power law characterized by a spectral tilt $q_a$, whose value and evolution remain under debate. Several simulations find an approximately scale-invariant spectrum with $q_a\simeq 1$~\cite{Buschmann:2021juv, Benabou:2024msj, Correia:2025nns}, while others report an increasing spectral index as the hierarchy between the Hubble radius and the string width grows, leading to an emission dominated in the infrared (IR) and with $q_a>1$~\cite{Gorghetto:2020qws, Kim:2024wku, Saikawa:2024bta}. The extrapolation of these results to cosmological scales is not yet settled. The resulting ALP population has a highly nonthermal momentum distribution covering many orders of magnitude. Even after most of the population becomes non-relativistic, ALPs in its high momentum tail suppress the growth of matter perturbations through free streaming~\cite{Amin:2022nlh, Liu:2024pjg, Harigaya:2025pox}, leading to observable signatures in the matter power spectrum and cosmic microwave background (CMB) lensing. Previous studies have investigated free streaming and isocurvature perturbations generated during postinflationary axion production~\cite{Gorghetto:2025uls, Chathirathas:2025aan, Dror:2026nfc}, while complementary CMB analyses have constrained coherently oscillating ultralight axions through their characteristic suppression of structure~\cite{Hlozek:2014lca, Lague:2026sbd}.

    In this work, we implement the nonthermal phase-space distribution of ALPs produced by cosmic strings in the Boltzmann code \class~\cite{Lesgourgues:2011rh}, accounting for its evolution at both the background and perturbative levels. To assess the uncertainty in the emission spectrum, we consider two benchmark scenarios with spectral tilt $q_a=1$ and $q_a=2$. We study how the free-streaming of ALPs depends on their mass and spectral tilt, and quantify its impact on the linear matter power spectrum and CMB lensing, beyond the usual treatment of ALPs as a cold DM component. We then perform a Markov Chain Monte Carlo (MCMC) analysis to constrain the ALP abundance, jointly with the six $\Lambda$CDM parameters, for different masses in the range $10^{-21} \, \mathrm{eV} \le m_a \le 10^{-12} \, \mathrm{eV}$. We combine CMB measurements from {\it Planck}~\cite{Planck:2018vyg}, the Atacama Cosmology Telescope (ACT) DR6~\cite{AtacamaCosmologyTelescope:2025blo} and the South Pole Telescope (SPT-3G) D1~\cite{SPT-3G:2025bzu}, together with baryon acoustic oscillation (BAO) measurements from the Dark Energy Spectroscopic Instrument (DESI) data release 2 (DR2)~\cite{DESI:2025zgx}. We consider different data cuts, including conservative cuts intended to minimize the potential impact of nonlinear corrections in the computation of theoretical predictions.

    The paper is organized as follows. In \cref{sec:ALP_production}, we describe the production of ALPs from a cosmic string network, with the resulting energy spectrum and phase-space distribution. In \cref{sec:pheno}, we discuss their cosmological phenomenology, including the background evolution, free-streaming, and their imprints on the matter power spectrum and CMB lensing. In \cref{sec:methods}, we describe the datasets and the methodology applied in our analysis, while in \cref{sec:results} we discuss the resulting constraints. We conclude in \cref{sec:conclusions}. We work in units $\hbar = c = 1$ and with Planck's mass $m_{\rm pl}$.

\section{Production of ALPs from cosmic string networks} \label{sec:ALP_production}

    The formation of a cosmic string network through the Kibble mechanism requires the underlying Peccei-Quinn (PQ) like global $U(1)$ symmetry, here $U(1)_{\rm PQ}$, to break after inflation~\cite{Kibble:1976sj,Kibble:1980mv}. This can occur if the symmetry remains unbroken during inflation or is subsequently restored through thermal effects during reheating or inflationary fluctuations~\cite{Turner:1990uz, Lyth:1992tw, Kofman:1995fi, Kasuya:1996ns, Kasuya:1997td}. The complex scalar field associated with the symmetry contains a massive radial excitation and a light angular mode, identified with the ALP. The resulting network of global strings loses energy predominantly through the emission of ALPs, with subdominant contributions from massive radial excitations and gravitational waves~\cite{Davis:1986xc, Harari:1987ht, Battye:1993jv}. As the Universe expands, the network evolves towards a scaling regime characterized by the Hubble scale, up to logarithmic corrections that depend on the ratio between the Hubble radius and the string width~\cite{Yamaguchi:1998gx, Yamaguchi:1999dy, Gorghetto:2018myk, Kawasaki:2018bzv, Vaquero:2018tib, Gorghetto:2021fsn, Hindmarsh:2021vih}. The scaling behavior is kept by the interplay between the stretching of long strings by cosmic expansion and their interactions, which lead to reconnections, loop formation, and the continuous energy loss through radiation.

    For the temperature independent ALP mass $m_a$ considered here, the explicit breaking of the global symmetry becomes dynamically relevant when the Hubble rate approaches $m_a$, leading to the formation of domain walls bounded by strings~\cite{Sikivie:1982qv, Vilenkin:1982ks}. For domain wall number $N_{\rm DW}=1$, each string is attached to a single wall, whose tension drives the collapse of the string-wall network shortly after its formation~\cite{Chang:1998tb, Hiramatsu:2010yn, Hiramatsu:2012gg}. For $N_{\rm DW}>1$, multiple walls are attached to each string, and the network can remain stable without additional explicit symmetry breaking~\cite{Sikivie:1982qv, Hiramatsu:2010yn, Hiramatsu:2012sc}, potentially leading to the domain wall problem~\cite{Zeldovich:1974uw, Lazarides:1982tw}. Sufficiently low tension domain walls may remain subdominant until the present epoch, with their tension constrained by CMB observations~\cite{Caloni:2026dyu}. We restrict our analysis to $N_{\rm DW}=1$ and consider ALP production during the scaling regime, neglecting the additional contribution during the collapse of the string-wall network.

    Numerical simulations of global string networks are challenging due to the large hierarchy between the string core width, set by the inverse radial mass $m_r^{-1}$, and the Hubble radius $H^{-1}$. A reliable simulation needs enough lattice points to faithfully resolve the string core, while covering a sufficiently large volume to capture the evolution of the network and minimize finite volume effects. This hierarchy is characterized by $\ln(m_r/H)$, with $m_r$ typically of order the symmetry breaking scale $f_a$. For $m_r\sim10^{14}\,\mathrm{GeV}$ and the ALP masses considered here, the hierarchy leads to $\ln(m_r/H)\sim80$--$100$ at the end of the scaling regime. Existing simulations explore values of this parameter of order $6$--$10$, depending on the numerical approach and resolution~\cite{Gorghetto:2018myk, Hindmarsh:2020pvl, Gorghetto:2020qws, Hindmarsh:2021vih, Gorghetto:2021fsn, Buschmann:2021juv, Baeza-Ballesteros:2023say, Benabou:2024mrz, Saikawa:2024bta}. Extrapolating these results to cosmological scales introduces uncertainties associated with the network evolution and the emitted ALP spectrum~\cite{Kaltschmidt:2025nkz}. 

\begin{figure}
        \centering
        \includegraphics[width=1\linewidth]{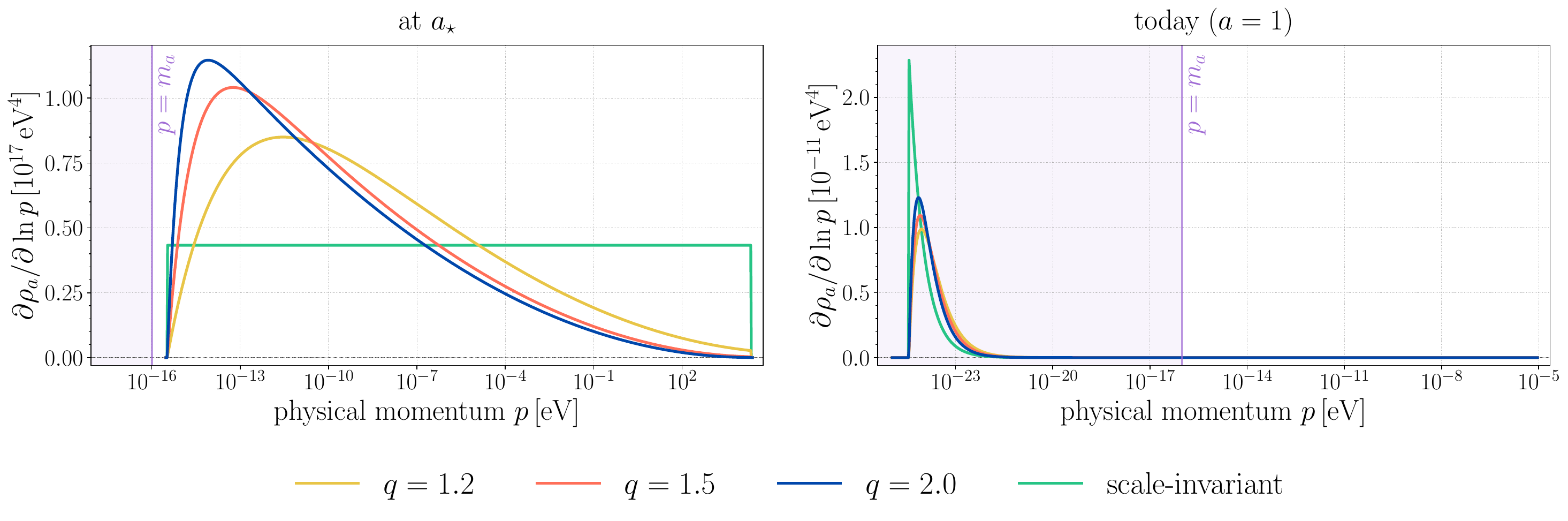}
        \caption{\textit{Left panel:} ALP energy spectrum $\partial \rho_a / \partial \ln p$ at the end of the scaling regime for different values of the tilt, including the scale-invariant benchmark $q_a = 1$. In this case, the energy density is uniformly distributed over logarithmic momentum intervals, while for $q_a > 1$ spectra favor lower momenta. \textit{Right panel:} Corresponding spectra today, illustrating the redshifting of the momentum distribution and a long tail at high momentum. For comparison, the spectra are normalized to have the same integrated energy density within each panel.}
        \label{fig: spectra and tilts}
    \end{figure}

    We model the cosmological ALP energy spectrum at a reference time $t_\star$ defined by $H(t_\star) \equiv H_\star = m_a/3$, shortly before the collapse of the string-wall network. The momentum distribution is highly nonthermal and extends over a broad range of momenta, from an IR scale $p \sim m_a$ to an effective ultraviolet (UV) scale $p_\star^{\rm max} \sim 10\sqrt{m_a f_a}$ corresponding to the axions released at the onset of the scaling regime and with momentum of order $m_r \sim  f_a$, redshifted until $H\sim m_a$. The precise shape and normalization of the distribution are uncertain and depend on logarithmic corrections, the evolution of the number of strings per Hubble volume $\xi$, and the spectral tilt $q_a$ of the instantaneous emission rate. The value and evolution of $q_a$ remain under debate, with $q_a = 1$ corresponding to a scale-invariant instantaneous emission and $q_a>1$ favoring emission into IR modes. To account for this uncertainty, we consider two benchmark prescriptions:
    \begin{itemize}
        \item $q_a >1.$ The ALP energy spectrum, obtained by integrating the instantaneous emission rate over the scaling regime, is given by~\cite{Gorghetto:2020qws, Gorghetto:2021fsn, Kim:2024wku, Kim:2024dtq, Chathirathas:2025aan}
        \begin{align}
            \frac{\partial \rho_a}{\partial p}(t_{\star}, p\mid q_a>1) =&\frac{8\xi_a \mu_a H^2_\star}{p}\Biggl\{\left[1-2\frac{\ln(p/\pmin)}{\ln_\star}\right]^2-\left(\frac{\pmin}{p}\right)^{q_a-1}\notag \\ 
            & +4\frac{1-2\frac{\ln(p/\pmin)}{\ln_\star}-\left(\frac{\pmin}{p}\right)^{q_a-1}}{(q_a-1)\ln_\star}+8\frac{1-\left(\frac{\pmin}{p}\right)^{q_a-1}}{(q_a-1)^2\ln_\star^2}\Biggr\} \, , \label{eq:spectrum_qgr1}
        \end{align}
        where $\ln_\star \equiv \ln(m_r/m_a)=\ln(m_r/3H_\star)$ gives the logarithmic hierarchy~\cite{Saikawa:2024bta}, $\xi_a \simeq 0.24 \ln_\star$ is the number of strings per Hubble volume as in~\cite{Gorghetto:2021fsn}, and $\mu_a=\pi f_a^2 \ln_\star$ is the effective string tension. We focus on the benchmark $q_a=2$.
        \item $q_a = 1.$ The energy spectrum for a scale-invariant emission is modeled phenomenologically as
        \begin{equation} \label{eq:spectrum_qeq1}
            \frac{\partial \rho_a}{\partial p}(t_\star, p \mid q_a =1)= \frac{A}{p} \, ,
        \end{equation}
        where the normalization $A$ depends on $f_a$, $\xi$, and other string network properties. \Cref{eq:spectrum_qeq1} is defined independently, since the expression in \cref{eq:spectrum_qgr1} does not provide a properly normalized scale-invariant spectrum in the limit $q_a\to 1$.
    \end{itemize}
    For both benchmarks, we adopt the momentum range $\pmin\leq p\leq p_\star^{\rm max}$, with $\pmin\equiv10 H_\star= 10 m_a/3$ and $p_\star^{\rm max}\equiv10\sqrt{f_am_a}$. Both spectra extend over many orders of magnitude in momentum, and show an approximately scale-invariant behavior. While the $q_a=1$ spectrum distributes energy uniformly over logarithmic momentum intervals, the $q_a>1$ spectra favor lower momenta. The corresponding spectra are shown in the left panel of \cref{fig: spectra and tilts}.

\section{Cosmological phenomenology} \label{sec:pheno}

    We explore the cosmological evolution of ALPs produced by cosmic strings, beginning with their background energy density and pressure. We then study the evolution of their perturbations, focusing on the damping of small scale fluctuations due to free-streaming, and the corresponding observational imprints on the matter and CMB anisotropies.

    \subsection{Background evolution} \label{sec:background}

        We first determine the unperturbed phase-space distribution $\mathcal{F}_0(p,\,t_\star)$ corresponding to the energy spectra in \cref{eq:spectrum_qgr1,eq:spectrum_qeq1}, following the collisionless evolution for $t > t_{\star}$. The ALP energy density  at $t_\star$ is given by\footnote{The factor $g/(2\pi)^3$, accounting for the internal degrees of freedom and phase-space normalization, is absorbed into the definition of $\mathcal{F}_0$.}
        \begin{equation}
            \rho_a(t_\star) = \int {\rm d}p \, \frac{\partial{\rho} }{\partial p } = 4\pi \int {\rm d}p \, p^2 \sqrt{p^2+m_a^2} \, \mathcal{F}_0(p, t_\star) \, .
        \end{equation}
        Using the two energy spectra, we obtain
        \begin{align}
            \mathcal F_0(p,t_\star) &= \frac{2\xi_a\mu_a H_\star^2}{\pi p^3\sqrt{p^2+m_a^2}}\Biggl\{\left[1-\frac{2\ln(p/p_\star^{\rm min})}{\ln_\star}\right]^2-\left(\frac{p_\star^{\rm min}}{p}\right)^{q_a-1} \notag \\
            &\quad + 4 \frac{1-2\frac{\ln(p/\pmin)}{\ln_\star}-\left(\frac{\pmin}{p}\right)^{q_a-1}}{(q_a-1)\ln_\star}+8 \frac{1-\left(\frac{\pmin}{p}\right)^{q_a-1}}{(q_a-1)^2\ln_\star^2}\Biggr\},
            &\quad \text{for }q_a>1 , \label{eq:distribution_qgr1} \\
            \mathcal F_0(p,t_\star) &= \frac{A}{4\pi p^3\sqrt{p^2+m_a^2}}, &\quad \text{for }q_a=1 \, . \label{eq:distribution_qeq1}
        \end{align}
        Both distributions vanish outside the momentum range $\pmin\le p \le p_\star^{\rm max}$.

        Neglecting further ALP production during the collapse of the string-wall network, we assume that the ALP population is collisionless after the collapse of the network. Introducing the comoving momentum $q=ap/a_\star$, the phase-space distribution is time independent, since $q$ accounts for the redshifting of the physical momentum. The background energy density and pressure can then be expressed as
        \begin{subequations}
            \label{eq:rhopres}
            \begin{align}
                \rho_a (\tilde{a}) &= \calN_0 \tilde{a}^{-4}  \int_{1}^{y_{\rm max}} \frac{{\rm d}y}{y} \, g(y) \left[ y^2+\left(\frac{\tilde a m_a}{\pmin}\right)^2 \right]^{1/2} \left[ y^2+\left(\frac{m_a}{\pmin}\right)^2 \right]^{-1/2} \, , \\[0.2cm]
                P_a (\tilde{a}) &= \frac{1}{3}\calN_0 \tilde{a}^{-4}  \int_{1}^{y_{\rm max}} {\rm d}y \, y\, g(y) \left\{\left[y^2+\left(\frac{\tilde a m_a}{\pmin}\right)^2\right] \left[ y^2+\left(\frac{m_a}{\pmin}\right)^2\right] \right\}^{-1/2} \, ,
            \end{align}
        \end{subequations}
        where $y \equiv p/ \pmin$ and $y_{\rm max} = 3\sqrt{f_a/m_a}$, we have defined the scale factor normalized to 1 at the time of network destruction, $\tilde a \equiv a(t)/a_\star=(1+z_\star)/(1+z)$, the normalization factor $\calN_0 \equiv \left\{8\xi_a\mu_a H_\star^2, \, A \right\} $ for $q_a >1$ or $=1$, respectively, and
        \begin{equation} \label{eq:gdef}
            g(y) \equiv \left\{
            \begin{array}{@{}c@{\qquad}c@{}}
                \begin{aligned}
                    \left(1-2\frac{\ln y}{\ln_\star}\right)^2 \!-\! y^{1-q_a} \!+\! 4 \,\frac{\displaystyle1-2\frac{\ln y}{\ln_\star}-y^{1-q_a}}{(q_a-1)\ln_\star} \!+\! \frac{8 \left(1-y^{1-q_a}\right)}{(q_a-1)^2\ln_\star^2}\,,
                \end{aligned}
                &  \hbox{for }q_a > 1\,, \\[4ex]
                1 \, , & \hbox{for }q_a =1\,.
            \end{array} \right.
        \end{equation}
        Since $\pmin = 10H_\star$, the integrals in \cref{eq:rhopres} at fixed $\tilde a$ and $q_a$, depend on the ratio $f_a/m_a$ through the upper integration limit and, for $q_a>1$, through the logarithmic hierarchy $\ln_\star = \ln(f_a/m_a)$.

        At the end of the scaling regime, ALPs are highly relativistic since their momenta satisfy $p > \pmin = 10H_\star$. As such, in this initial regime ALPs behave as radiation, with an equation of state $w_a \equiv P_a/\rho_a \simeq 1/3$ and the energy density scaling as $a^{-4}$. As the Universe expands, the ALP momenta redshift and an increasingly large fraction of particles becomes non-relativistic. Eventually, ALPs behave as a DM component, with $w_a \simeq 0$ and $\rho_a \propto a^{-3}$.

\begin{figure}
            \centering
            \includegraphics[width=0.8\linewidth]{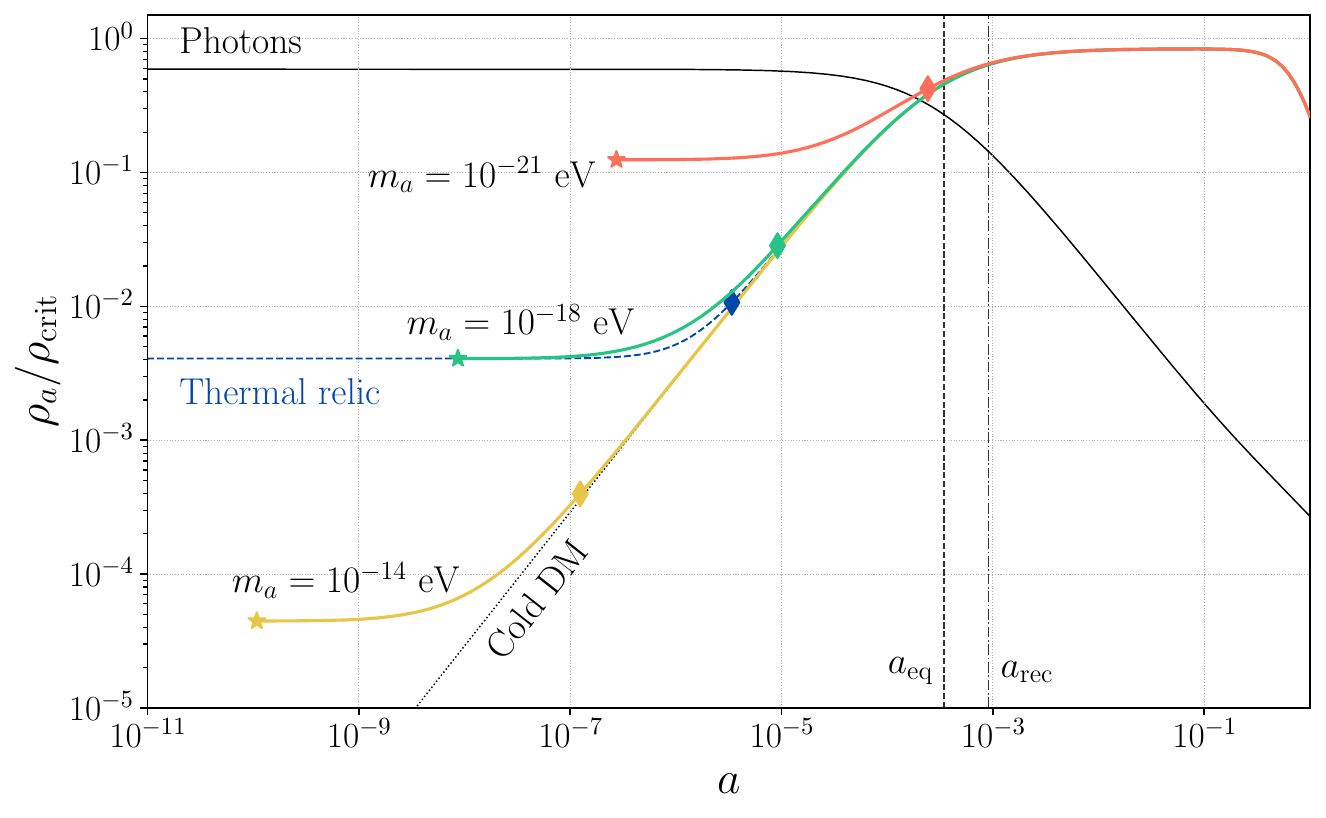}
            \caption{Evolution of the ALP density parameter $\Omega_a \equiv \rho_a/\rho_{\rm crit}$ for different ALP masses, starting from the end of the scaling regime (marked with stars). For each mass, the distribution is normalized so that the present-day abundance is $\Omega_a=0.26$. For each curve, we also highlight the time of non-relativistic transition (diamonds), defined as the epoch when 90\% of the ALP energy density is carried by non-relativistic modes. Networks associated with lighter ALPs are destroyed at later times, so that the ALPs undergo less redshifting and become non-relativistic at later times. For comparison, we show the density parameters of a cold dark matter (CDM) component (dashed black line) and photons (solid black line). We also include an illustrative thermal relic with a Bose--Einstein distribution, with the same relativistic density of ALPs with mass $10^{-18}$\,eV and the same non relativistic density of CDM. The thermal distribution shows a sharper transition between relativistic and non-relativistic scaling than the nonthermal ALP spectra. Vertical lines mark matter-radiation equality and recombination.}
            \label{fig: thermo mass}
        \end{figure}

        Since $H(t_\star) = m_a/3$, the relation between the normalized scale factor $\tilde a$ and redshift $z$ depends on the ALP mass. We derive the corresponding photon temperature $T_\star \equiv T(z_\star)$ by solving\footnote{We assume radiation domination and neglect the ALP contribution to the total energy density at $t_\star$. We verify a posteriori that these assumptions hold in the region of parameter considered in our analysis, see e.g.~\cref{fig: thermo mass}.}
        \begin{equation}
            H_\star = 1.66 g_{*}(T_\star)^{1/2} \frac{T_\star^2}{m_\mathrm{pl}} =\frac{m_a}{3} \, ,
        \end{equation}
        where $T$ denotes the photon temperature, $g_*(T)$ is the effective number of relativistic degrees of freedom for energy density,\footnote{Note that the $*$ subscript in $g_*$ is not related to $\star$ subscript. We use the standard notation for this quantity assuming it does not generate confusion.} and $m_{\rm pl}$ is Planck's mass. The value of $1+z_\star$ can then be found using entropy conservation in the form $g_{*s}^{1/3}aT = \mathrm{const.}$, where $g_{*s}(T)$ is the effective number of relativistic degrees of freedom contributing to the entropy density. If the temperature dependence in $g_*$ and $g_{*s}$ were neglected, both $T_\star$ and $1+z_\star$ would be simply proportional to $m_a^{1/2}$. Since in practice $g_*$ and $g_{*s}$ vary very slowly with temperature, going from $\sim 100$ at hundreds of GeV to $\sim 3$ below MeV, we still expect the scaling of $T_\star$ and $1+z_\star$ with the ALP mass to be very close to $m_a^{1/2}$. Using $m_a = 10^{-21}$\,eV, we express the temperature and redshift at the end of the scaling regime as
        \begin{subequations}
            \begin{align}
                T_\star &= 1.16\times 10^{-6}{\rm\,GeV} \, c_T(m_a)\sqrt{\frac{m_a}{10^{-21}\,{\rm eV}}}\,,\\[0.3cm]
                1+z_\star &= 4.92\times 10^6 \, c_z(m_a)\sqrt{\frac{m_a}{10^{-21}\,{\rm eV}}}\,,
            \end{align}
        \end{subequations}
        where the correction functions $c_T(m_a)$ and $c_z(m_a)$ are normalized to unity at $m_a=10^{-21}$\,eV, and account for the temperature dependence of the effective relativistic degrees of freedom,
        \begin{equation}
            c_T(m_a) = \left[\frac{3.36}{g_*(T_\star)}\right]^{1/4},
            \qquad
            c_z(m_a) = c_T(m_a) \left[\frac{g_{*s}(T_\star)}{3.91}\right]^{1/3}\,.
        \end{equation}
        Since the network is destroyed later for smaller ALP masses, the spectrum at any given time will be ``hotter'' for smaller values of $m_a$. This behavior is shown in figure~\ref{fig: thermo mass}, where we choose $q_a = 1$, we vary $m_a $ in eq.~\eqref{eq:distribution_qeq1}  and rescale the distribution function in order to have $\Omega_a$ equal to a reference value, that we take to be $\Omega_a=0.26$. We can appreciate how the change in the mass roughly corresponds to a shift in the momentum distribution.

\begin{figure}
            \centering
            \includegraphics[width=0.49\linewidth]{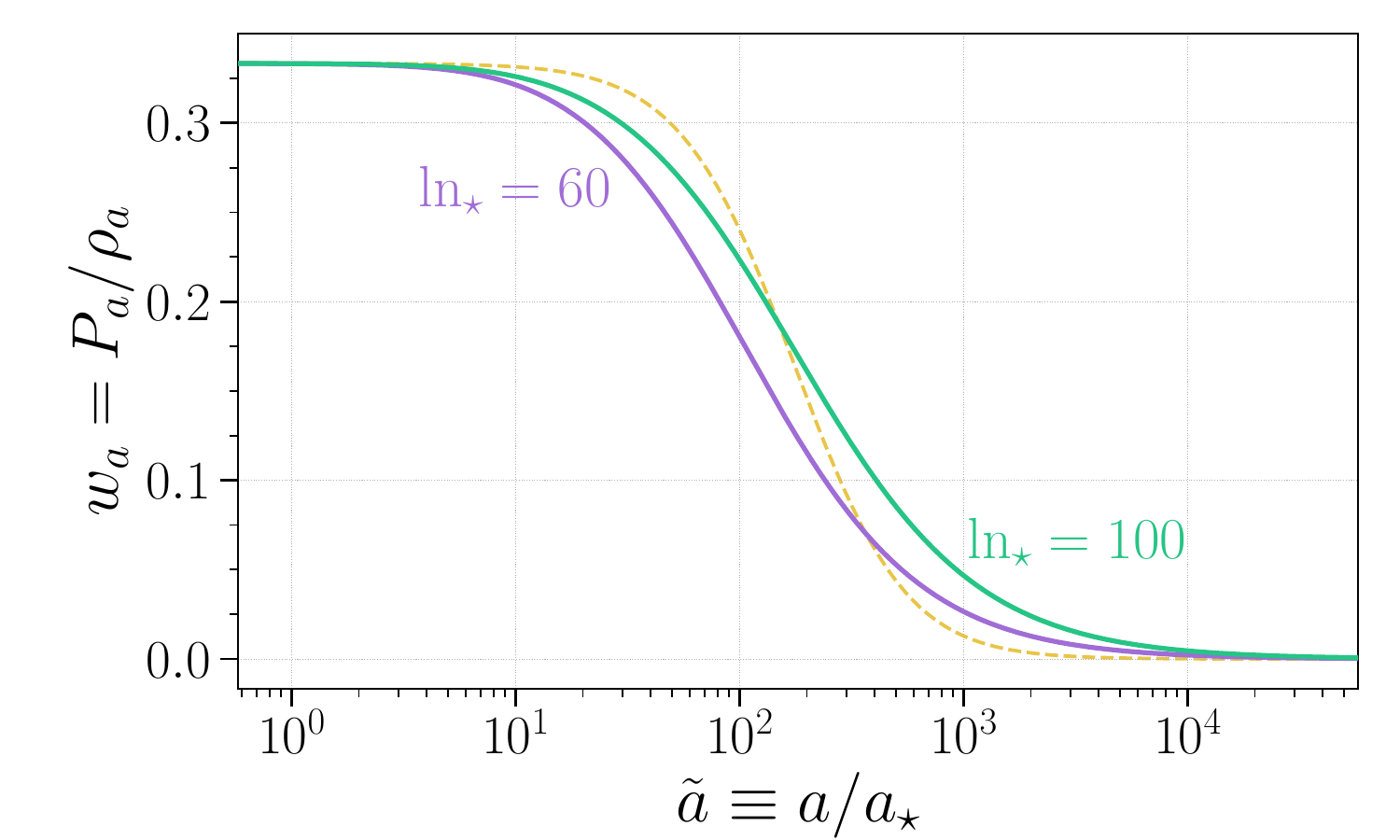}
            \includegraphics[width=0.49\linewidth]{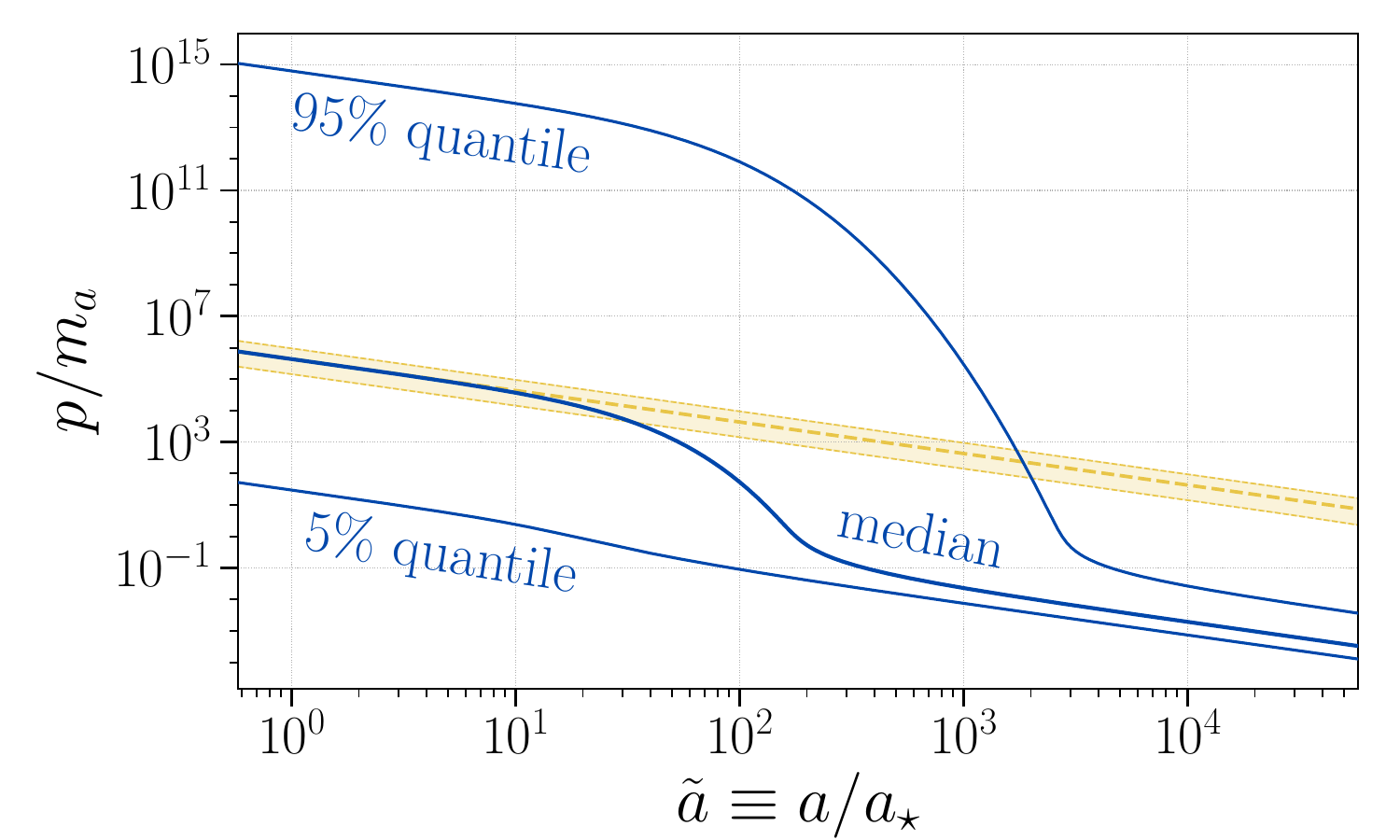}
            \caption{\textit{Left panel:} Evolution of the ALPs equation-of-state parameter $w_a \equiv P_a /\rho_a$ as a function of the normalized scale factor $\tilde a\equiv a/a_\star$, where $a_\star$ is the scale factor at the time of network destruction. The violet and green curves correspond to $\ln_\star = 60$ and $\ln_\star = 100$, respectively. For comparison, the dashed yellow curve shows a thermal ALP with the same ratio of the mean momentum to its mass, $\langle p \rangle/m \sim 300$. \textit{Right panel:} Evolution of the physical momenta quantiles enclosing 5\%, 50\% (median), and 95\% of the total ALP energy density. Solid blue curves correspond to ALPs produced from cosmic strings, while dashed yellow curves are for a thermal ALP whose initial median momentum matches that of the nonthermal distribution at $\tilde a=1$. The broad momentum distribution of ALPs from cosmic strings leads to an extended transition to nonrelativistic behavior, with a high-momentum tail persisting even after the median momentum falls below $m_a$.}
            \label{fig:wa_vs_a}
        \end{figure}

        In \cref{fig:wa_vs_a} (left panel), we show the evolution of the ALP equation-of-state parameter $w_a = P_a/\rho_a$. Since the normalization factors cancel out in the pressure-to-density ratio, the evolution, when expressed in terms of $\tilde a$ as a time variable, only depends on $\ln_\star$ and on the ratio $\pmin/m_a$. The latter is fixed to (10/3) in our analysis, while we show $w_a$ for $\ln_\star = 60,\,100$ in the plot (orange and blue curves). For comparison, we also show $w_a$ for a thermal ALP with a temperature $T(a_\star) \simeq 100 m_a$ (dashed green curve). The value of the temperature is chosen so that the average momentum-to-mass ratio $\langle p \rangle /m_a$ is the same for the thermal and the string-produced ALP with $\ln_\star=100$. It is worth noting how for string-produced ALPs the transition from the ultrarelativistic ($w_a\simeq1/3$) to non-relativistic ($w_a\simeq 0$) regimes extends over a longer period of time, as compared to thermal particles. This is due to the large width of the energy spectrum in the former case.
        
        In order to better appreciate how the width of the momentum distribution changes over time, we compute quantiles of this distribution. In \cref{fig:wa_vs_a} (right panel), we show the physical momenta $p$ enclosing the lower 5\%, 50\% (i.e. the median) and 95\% of the total ALP energy density, as a function of $\tilde{a}$. For comparison, we also show the same quantiles for a thermal ALP. The temperature of the latter is chosen so that the median of the two distributions is the same at $\tilde{a}=1$.

    \subsection{Free-streaming of ALPs from string networks}

        A distinctive feature of ALPs produced by cosmic string networks is their highly nonthermal momentum distribution, spanning many orders of magnitude. For this, a large fraction of the ALP population can remain relativistic long after the end of the scaling regime. The residual velocities allow ALPs to free-stream out of potential wells, suppressing the clustering of matter, similarly to massive relic neutrinos. This effect is particularly relevant for smaller ALP masses, for which the transition to non-relativistic behavior occurs later.

        We define an effective comoving free-streaming horizon,
        \begin{equation} \label{eq:fs_horizon}
            r_{\rm EFF}(\eta) \equiv \int_{\eta_\star}^{\eta}{\rm d}\eta'\, v_{\rm rms}(\eta') \, ,
        \end{equation}
        where $\eta$ is conformal time and $v_{\rm rms}$ the root-mean-square (rms) particle speed. The corresponding wavenumber $k_{\rm EFF}\equiv 2\pi/r_{\rm EFF}$ provides a characteristic free-streaming scale. ALPs from cosmic strings show a slower decrease of their velocity dispersion due to the presence of a high-momentum tail. For $q_a=1$, the initial phase-space distribution behaves approximately as $\mathcal{F}_0(p)\propto p^{-4}$ and the velocity dispersion is
        \begin{equation}
            v_{\rm rms}^2(\tilde a) = \frac{\displaystyle\int_{\pmin}^{p_\star^{\rm max}}{\rm d}p\,\frac{p^4}{p^2+\tilde a^2m_a^2}\mathcal{F}_0(p,t_\star)}{\displaystyle\int_{\pmin}^{p_\star^{\rm max}}{\rm d}p\,p^2\mathcal{F}_0(p,t_\star)}\,.
        \end{equation}
        In the intermediate regime $\pmin\ll\tilde a m_a\ll p_\star^{\rm max}$, implying $v_{\rm rms}\propto a^{-1/2}$ until the highest momentum modes become non-relativistic.

        To estimate the resulting free-streaming scale, we introduce a characteristic epoch $\eta_v$ at which the lowest momentum modes become non-relativistic, corresponding to $a_v/a_\star\sim\pmin/m_a$. Neglecting the contribution after matter-radiation equality with conformal time $\eta_{\rm eq}$, we obtain
        \begin{equation}
            r_{\rm EFF}(\eta_{\rm eq}) \sim \int_{\eta_\star}^{\eta_v}{\rm d}\eta' + \sqrt{\eta_v}\int_{\eta_v}^{\eta_{\rm eq}}\frac{{\rm d}\eta'}{\sqrt{\eta'}} = 2\sqrt{\eta_v\eta_{\rm eq}} - \eta_v-\eta_\star\,.
        \end{equation}
        For $\eta_{\rm eq}\gg\eta_v$, the effective horizon scales as $r_{\rm EFF}\propto\sqrt{\eta_v}$. Since during radiation domination $a_v\propto a_\star\propto H_\star^{-1/2}\propto m_a^{-1/2}$, we obtain the approximate scaling $k_{\rm EFF}\propto m_a^{1/4}$, up to corrections in the evolution of the relativistic degrees of freedom.

        \begin{table}[t]
            \centering
            \small
            \renewcommand{\arraystretch}{1.4} 
            \begin{tabular}{l c c c c c c}
                \hline\hline
                $m_a\,[\mathrm{eV}]$ & $10^{-12}$ & $10^{-14}$ & $10^{-16}$ & $10^{-18}$ & $10^{-20}$ & $10^{-21}$\\
                \hline
                $k_{\rm EFF}\,[\mathrm{Mpc}^{-1}]$ & $4.6$ & $1.4$ & $0.42$ & $0.13$ & $0.04$ & $0.02$\\
                $k_{\rm pheno}\,[\mathrm{Mpc}^{-1}]$ & -- & $1.3$ & $0.43$ & $0.15$ & $0.05$ & $0.03$\\
                $k_{\rm FS}\,[\mathrm{Mpc}^{-1}]$ & $1.2\times10^3$ & $1.0\times10^2$ & $9.4$ & $1.3$ & $0.15$ & $0.06$\\
                \hline\hline
            \end{tabular} 
            \caption{Characteristic free-streaming wavenumbers $ k_{\rm EFF}$, $k_{\mathrm{pheno}}$ and $k_{\rm FS}$, in units of $\mathrm{Mpc}^{-1}$, for different axion masses $m_a$, with $q_a = 1$, rounded to two significant figures. $k_{\rm pheno}$ does not appear for the largest mass, as a suppression of $1\%$ of the linear matter power spectrum is never actually reached within the scales we inspect.  All three of these quantities would be in the same ballpark in the case of standard neutrinos, while it is evident the different scaling of $k_{\rm FS}\propto \sqrt{m_a}$, which does not provide a good description of the true free-streaming scale, with respect to the other two.  }
            \label{tab:kfs_keff masses}
        \end{table}

        The numerical values of $k_{\rm EFF}$ for the $q_a=1$ benchmark are given in \cref{tab:kfs_keff masses}. Their mass dependence is approximately described by
        \begin{equation}\label{scaling law}
            k_{\rm EFF}(m_a) \simeq 0.13\left(\frac{m_a}{10^{-18} \, \text{eV}}\right)^{1/4} \, \text{Mpc}^{-1}\,.
        \end{equation}
        At the lower end of the mass range, $k_{\rm EFF}$ lies within the range of wavenumbers probed by CMB lensing, approximately $10^{-2} \lesssim k\,{\rm Mpc} \lesssim 0.5$~\cite{Green:2021xzn}. To assess the correspondence between this effective scale and the suppression of matter perturbations, we introduce a phenomenological wavenumber $k_{\rm pheno}$, defined as the scale at which the linear matter power spectrum is suppressed by $1\%$ relative to a cosmology in which the ALP component is replaced by CDM. For this comparison, ALPs constitute $1\%$ of the total DM abundance, with all other cosmological parameters been fixed. As shown in \cref{tab:kfs_keff masses}, the resulting values of $k_{\rm pheno}$ follow those of $k_{\rm EFF}$ over the mass range considered. For comparison, we also consider the conventional instantaneous free-streaming wavenumber, $k_{\rm FS} = \sqrt{3/2}\,a H(a) /v_{\rm rms}(a)$, which characterizes the instantaneous competition between particle streaming and gravitational clustering. This quantity scales as $k_{\rm FS} \propto a_\text{NR}^{-1} \propto m_a^{1/2}$, when evaluated near the non-relativistic transition during radiation domination. Unlike $k_{\rm EFF}$, however, it does not account for the cumulative particle displacements after this transition. The corresponding values differ from both $k_{\rm EFF}$ and $k_{\rm pheno}$ for large ALP masses.

        \begin{figure}[t]
            \centering
            \includegraphics[width=0.8\linewidth]{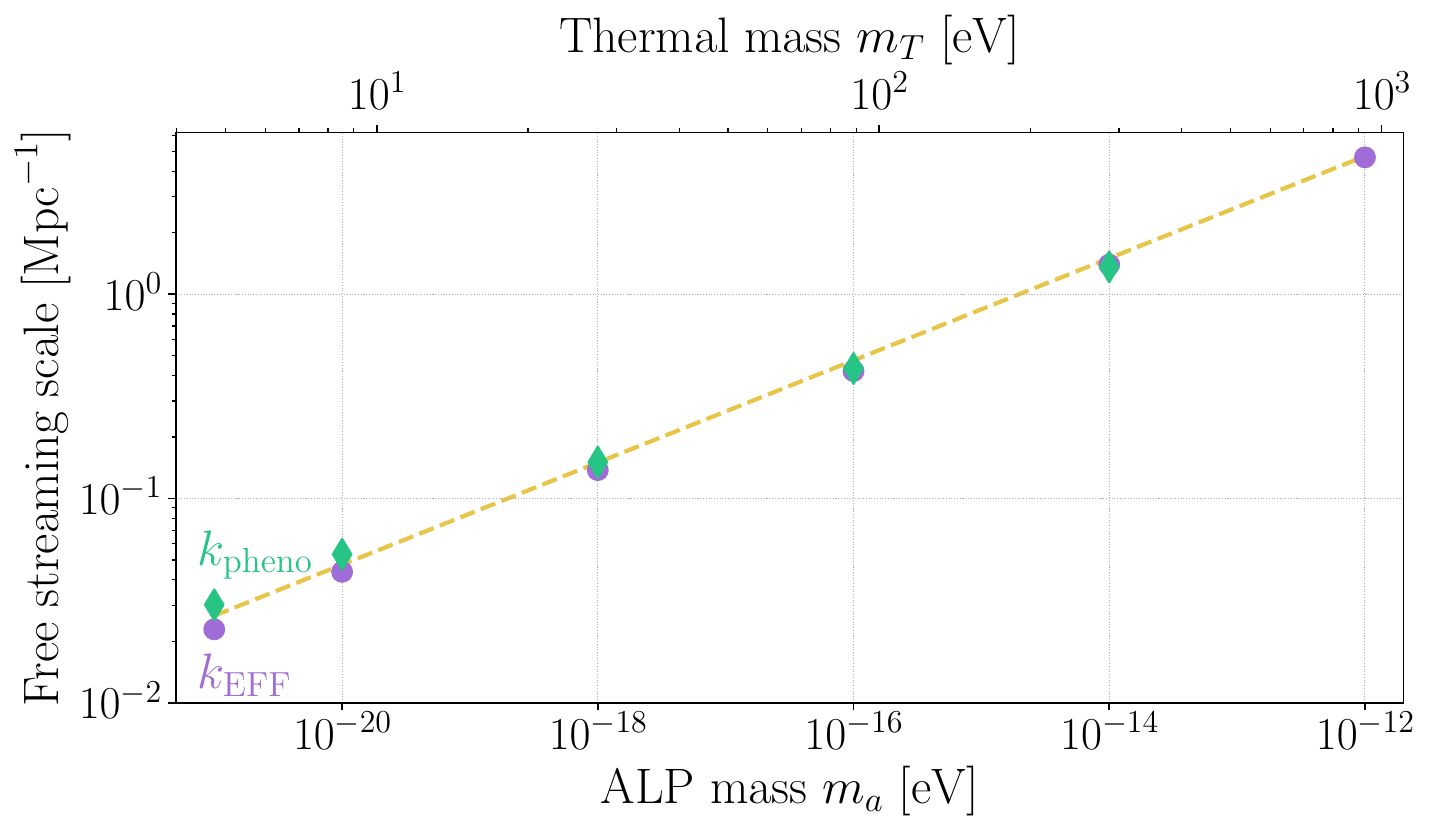}
            \caption{Comparison between the effective free-streaming wavenumber $k_{\rm EFF}$ and the phenomenological suppression scale $k_{\rm pheno}$ for ALP masses in the range $10^{-21}\,{\rm eV}\leq m_a\leq10^{-12}$\,eV. The horizontal axes show the ALP mass $m_a$ and the corresponding effective thermal mass $m_T$, related by \cref{thermal mass relation}. The dashed yellow line shows the scaling relation in \cref{scaling law}.}
            \label{fig:thermal_mass}
        \end{figure}

        Finally, we use the effective free-streaming horizon to map the ALP mass with an effective thermal mass $m_T$, defined as the mass of a thermal relic with a Fermi--Dirac distribution and a present temperature $T_0 = 1.9\,K$, so that the same free-streaming horizon as the corresponding ALP population is recovered. This gives
        \begin{equation} \label{thermal mass relation}
            \frac{m_T}{\rm eV} = 30\left(\frac{m_a}{10^{-18}\,{\rm eV}}\right)^{1/4}\,.
        \end{equation} 
        This correspondence is kinematical and concerns the estimate of the free-streaming horizon. It should not be used to translate cosmological constraints on thermal relic masses directly into bounds on $m_a$, since these also depend on the relic abundance and the shape of the momentum distribution. A related mapping between ultralight axions and thermal warm DM, based on the characteristic scale of structure suppression, was discussed in~\cite{Marsh:2015xka}. In \cref{fig:thermal_mass}, we compare $k_{\rm EFF}$ and $k_{\rm pheno}$ as functions of both $m_a$ and the corresponding effective thermal mass $m_T$.

    \subsection{Matter perturbations}

        We investigate the effects of string-produced ALPs on the linear matter power spectrum. We consider two sets of cosmological models in which ALPs produced from cosmic strings constitute either all of the DM, or a fraction equal to 1\%. In both cases, we assume one massive, with $ m_\nu = 0.06$\,eV, and two massless neutrino species, with standard thermal history, and total $N_{\rm eff}=3.044$. We consider ALP masses in the range $10^{-21}-10^{-12}$\,eV, and the two spectral benchmarks $q_a = 1$ and $q_a = 2$. All remaining cosmological parameters are fixed to their {\it Planck} best fit values.

\begin{figure}
            \centering
            \includegraphics[width=1\linewidth]{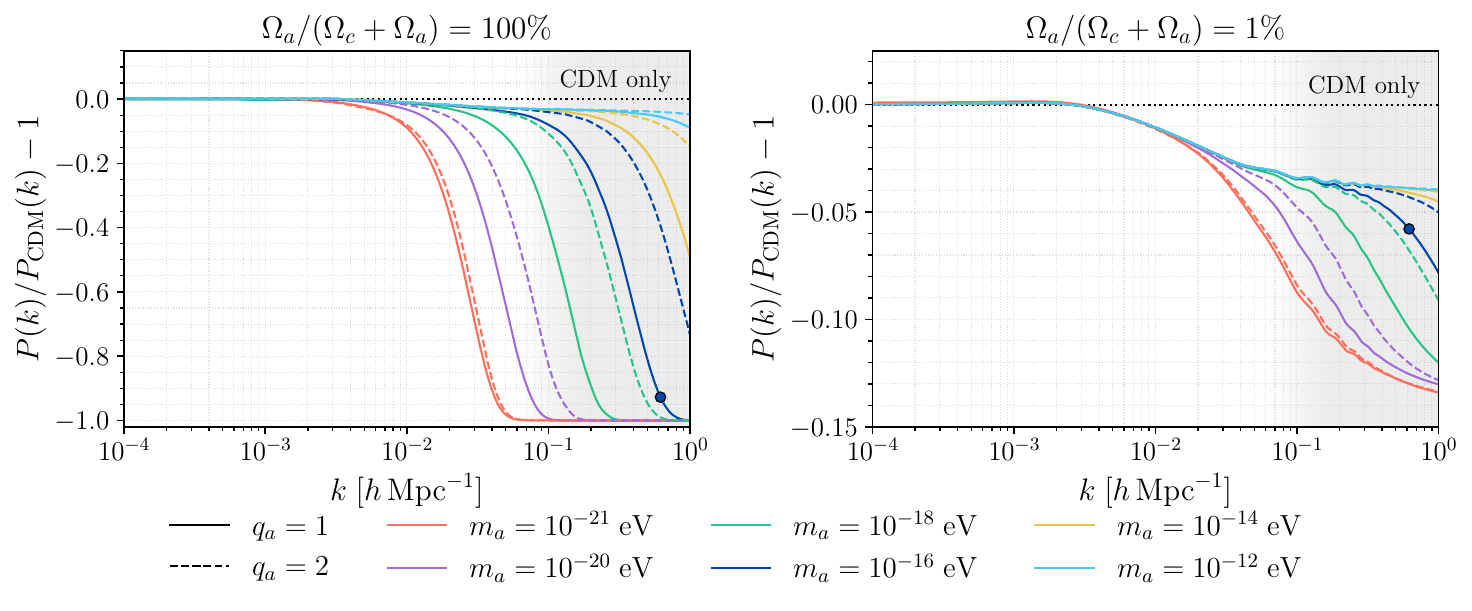}
            \caption{Fractional difference in the linear matter power spectrum, $P(k)/P_{\rm CDM}(k)-1$, relative to a reference $\Lambda$CDM cosmology with massless neutrinos. \textit{Left panel:} ALPs constitute the entirety of DM. \textit{Right panel:} ALPs constitute $1\%$ of the DM abundance. All models include one massive neutrino with $m_\nu=0.06\,\mathrm{eV}$ and two massless neutrinos. Colors indicate the ALP mass, while solid and dashed lines correspond to $q_a=1$ and $q_a=2$, respectively. The suppression associated with massive neutrinos is common to the ALP models, while the additional suppression from ALP free streaming depends on the ALP mass and spectral tilt. The dotted black line indicates the reference cosmology, and the gray shaded regions mark mildly nonlinear scales. Note the different vertical scales in the two panels.} 
            \label{fig: pk neutrinos}
        \end{figure}

        In \cref{fig: pk neutrinos}, we compare the resulting matter power spectra with a reference $\Lambda$CDM cosmology containing only cold DM and massless neutrinos, with the same $N_{\rm eff}$. Since massless neutrinos contribute negligibly to the matter density at late times, the reference model contains no massive free-streaming matter component. The comparison allows to disentangle the suppression of the matter power spectrum induced by massive neutrinos from that associated with the free-streaming of ALPs. The neutrino suppression in the spectra is common to the models considered, while the ALP free streaming introduces a new feature whose characteristic wavenumber depends on the ALP mass and spectral tilt. Sufficiently massive ALPs, with masses $\gtrsim 10^{-14}$\,eV, become essentially indistinguishable from CDM over the linear scales considered, even when they account for the entire DM abundance. For smaller masses $m_a \lesssim 10^{-16}$\,eV, the suppression is increasingly more evident. Increasing the spectral tilt favors lower momentum modes, shifting the onset of ALP free-streaming suppression towards larger wavenumbers.

        For comparison, the suppression coming from a subdominant free-streaming component can be estimated using the result derived for massive neutrinos~\cite{Hu:1997mj},
        \begin{equation} \label{wu et al}
            1-\frac{P(a,k)^X}{P(a,k)^0}\simeq 8f_X\,,
        \end{equation} 
        where $f_X \equiv \Omega_X/\Omega_m$ is the fraction of the total matter density contributed by the free-streaming species $X$, which we account for either neutrinos or ALPs, and ${P(a,k)^0}$ is the matter power spectrum computed in a CDM only cosmology. This approximation is valid for small fractions $f_X\lesssim 0.05$ and wavenumbers well above the suppression scale. For $\sum m_\nu=0.06$\,eV, the neutrino fraction is approximately $f_\nu\simeq0.0045$, corresponding to a suppression of order $4\%$. Similarly, an ALP component accounting for 1\% of the DM contributes an additional suppression of about 8\%. These estimates are indeed consistent with the approximate amplitudes in the right panel of \cref{fig: pk neutrinos}.

    \subsection{CMB lensing signatures} \label{subsec:CMB_spectra}

    \begin{figure}[t]
        \centering
        \includegraphics[width=0.8\linewidth]{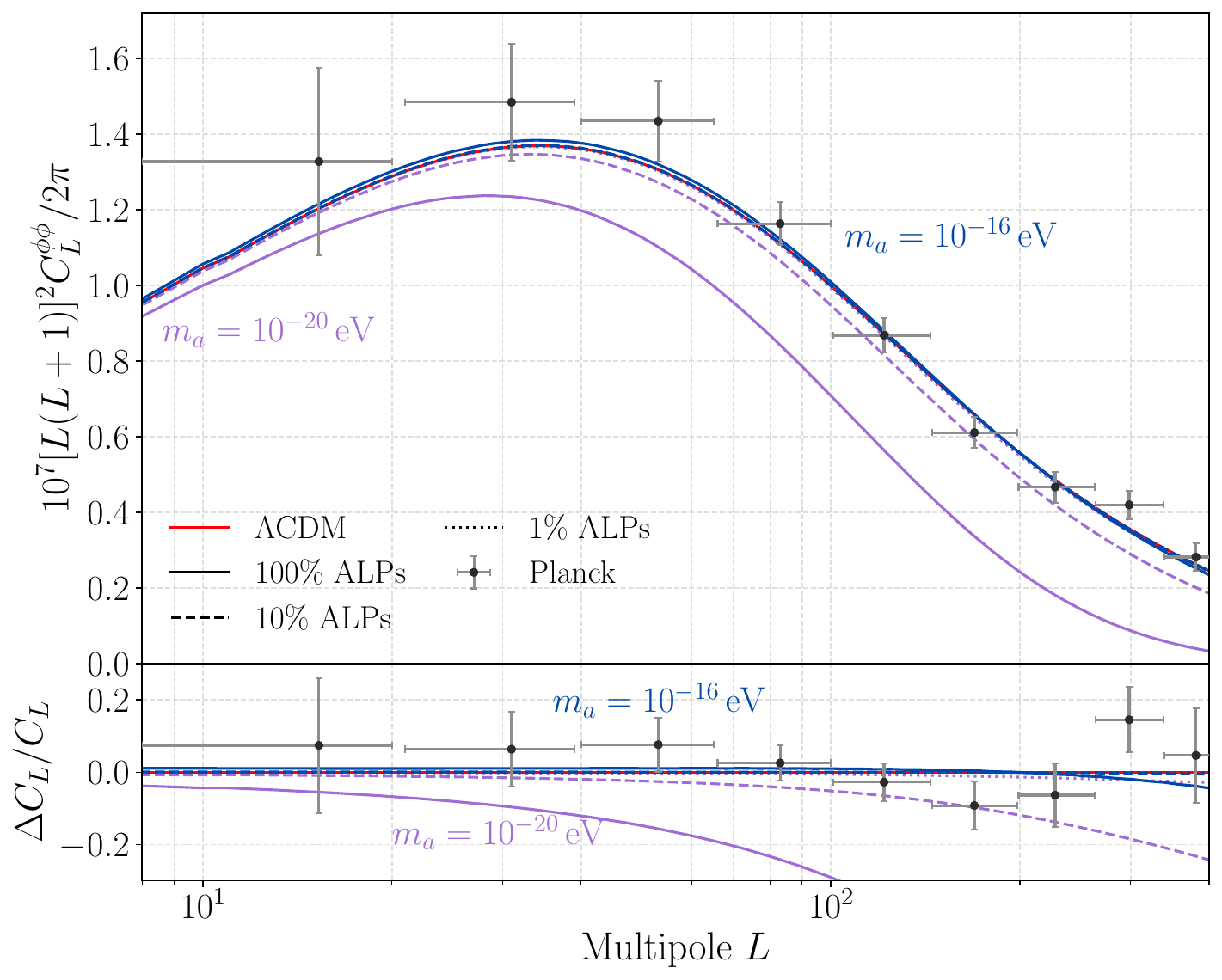}
        \caption{CMB lensing potential power spectrum $C_L^{\phi\phi}$ for different ALP abundances. The black line shows the $\Lambda$CDM prediction, with the six cosmological parameters fixed to their posterior mean values obtained from SPA3k+B+L dataset, see \cref{tab: lcdm}. Solid, dashed and dotted lines correspond to models in which $100\%, 10\% $, $1\%$ of the CDM is replaced by ALPs with $q_a= 1$, while keeping the other cosmological parameters fixed. Crosses show the {\it Planck} 2018 lensing measurements with their uncertainties. For lighter ALPs, free-streaming suppresses matter clustering, reducing the lensing power at higher $\ell$.}
        \label{fig: PP}
    \end{figure}

    We discuss the impact of ALPs produced by cosmic strings on the CMB lensing power spectrum, comparing our predictions with {\it Planck} 2018 lensing data~\cite{Carron:2022eyg}. This provides useful intuition for the constraints derived in the following section. Specifically, in \cref{fig: PP}  we show CMB lensing spectra computed by fixing the $\Lambda$CDM parameters to the mean of the posteriors of the run SPA3k+B+L, (more detail on the runs in the \cref{sec:results}, and in \cref{appendice: full triangle plots}) with three degenerate neutrinos of total mass $\sum m_\nu = 0.06\,\mathrm{eV}$, and compare them with the spectra obtained by replacing either 100\%,  $10\%$ or 1\%  of the CDM with ALPs. 

    For the lightest ALP masses ($m_a \lesssim 10^{-20} \, \mathrm{eV}$), we observe a suppression of power, starting at multipoles $L \gtrsim 100$, in the lensing power spectrum. This range lies in the window probed by CMB measurements~\cite{Green:2021xzn, Gerbino:2022nvz}. This suppression is proportional to the ALP fraction. The drop in the lensing power spectrum is a direct consequence of the ALP free-streaming (depending on the mass), while the impact of varying the tilt is subdominant. For a larger mass, the effect of free-streaming is pushed to smaller scales, away from the linear window.

\section{Methodology} \label{sec:methods}

    We perform a Markov Chain Monte Carlo (MCMC) analysis with \texttt{Cobaya}~\cite{2019ascl.soft10019T,Torrado:2020dgo}, interfaced with our modified version of \texttt{CLASS}\footnote{The modified version of \texttt{CLASS} and the code that supports the findings of this work is not publicly available, but it can be obtained from the authors upon reasonable request.}~\cite{Lesgourgues:2011re,Blas:2011rf}, to constrain the cosmological abundance of ALPs produced by cosmic strings. We consider four combinations of current state-of-the-art cosmological datasets:
    \begin{itemize}
        \item \textbf{SPA3k+B+L:} We combine temperature and polarization data from {\it Planck}, ACT DR6 and SPT-3G D1 with DESI DR2 BAO data~\cite{DESI:2025zgx}. For {\it Planck}, we use temperature data from the \texttt{Commander} likelihood~\cite{Planck:2018vyg,Planck:2019nip} and $E$-mode polarization data from \texttt{SRoll2}~\cite{Pagano:2019tci,Delouis:2019bub} at large angular scales ($2 \le \ell \le 29$), complemented at high multipoles by the \texttt{Plik\_lite} likelihood. Following ref.~\cite{AtacamaCosmologyTelescope:2025blo}, we restrict {\it Planck} data at multipoles $\ell_{\rm max} = 1000$ in temperature and $\ell_{\rm max} = 600$ in polarization to reduce overlap with ACT DR6. We use the \texttt{ACT-lite}\footnote{\url{https://github.com/ACTCollaboration/DR6-ACT-lite}} likelihood for ACT DR6~\cite{AtacamaCosmologyTelescope:2025blo}, restricted to $\ell_{\rm max} = 3000$ in both temperature and polarization instead of the nominal range $600\le \ell \le 6500$. We neglect correlations between the {\it Planck} and ACT datasets in the range where they overlap~\cite{AtacamaCosmologyTelescope:2025blo}. For SPT-3G D1~\cite{SPT-3G:2025bzu, Balkenhol:2024sbv}, we use the \texttt{SPT-lite}\footnote{\url{https://github.com/SouthPoleTelescope/spt_candl_data}} likelihood, restricted to multipoles from $\ell_\textrm{min} = 400$ to $\ell_\textrm{max} = 3000$ in both temperature and polarization. We assume no correlations between SPT and the other datasets, given the small overlap in their observed regions~\cite{SPT-3G:2025bzu}. In addition, we include the {\it Planck} PR4 lensing reconstruction likelihood~\cite{Carron:2022eyg}, restricted to $L_{\rm max}= 400$.\footnote{\url{https://github.com/carronj/planck_PR4_lensing}}. We refer to this as our baseline dataset.

        \item \textbf{SPA5k+B+L:} Same as SPA3k+B+L, but with ACT temperature and polarization spectra extended to $\ell_\text{max} = 5000$ and with SPT extended to its full multipole coverage ($\ell_\text{max}=3000$ for temperature and $\ell_\text{max}=4000$ for polarization). 

        \item \textbf{SPA3k+B:} Same as SPA3k+B+L, but without the inclusion of {\it Planck} PR4 lensing reconstruction likelihood. 

        \item \textbf{SPA5k+B:} Same as SPA5k+B+L, but without the inclusion of {\it Planck} PR4 lensing reconstruction likelihood.   
    \end{itemize}

    The precision of current CMB experiments, especially those that probe small angular scales, requires careful treatment of nonlinear corrections to the evolution of matter perturbations. These corrections affect CMB observables through their impact on weak gravitational lensing, modifying both the lensed temperature and polarization power spectra and the reconstructed lensing potential power spectrum. Nonlinear corrections are typically computed using prescriptions such as \texttt{Halofit}~\cite{Smith:2002dz,Bird:2011rb,Takahashi:2012em} and \texttt{HMcode}~\cite{Mead:2015yca,Mead:2016zqy,Mead:2020vgs} and are calibrated on $\Lambda$CDM. As such, support for extended models remains limited, raising concern when constraining cosmological models with interacting dark sectors~\cite{Trendafilova:2025dce} and ultralight axion models~\cite{Gaughan:2026xrv}. In the absence of a reliable prescription for computing nonlinear corrections in our ALP model, we compute the CMB observables using linear matter perturbations and restricting our analysis to the datasets and scales that are less sensitive to nonlinear evolution. The SPA3k+B dataset combination should be regarded as the most conservative choice. We also consider less conservative combinations that include the lensing reconstruction likelihood from \textit{Planck},\footnote{The authors of ref.~\cite{Trendafilova:2025dce} find that, within $\Lambda$CDM, neglecting nonlinear corrections induces shifts in the inferred parameters reaching approximately $0.2\sigma$ and $0.6\sigma$ when including lensing reconstruction data from {\it Planck} or SPT-3G 2018 respectively.} extend the multipole range of the temperature and polarization spectra to $\ell_\mathrm{max} = 5000$, or both. Throughout the analysis, we neglect nonlinear corrections to the matter power spectrum when computing the theoretical CMB observables. We anticipate that our constraints are largely insensitive to extending the primary CMB analysis from $\ell_{\max}=3000$ to $5000$, as well as to the inclusion of the {\it Planck} lensing reconstruction likelihood. These consistency checks provide additional reassurance that uncertainties associated with nonlinear corrections are unlikely to significantly affect our conclusions. We clarify that also for the cases of $\Lambda$CDM, the same datasets and the same hypotheses on the nonlinear conditions are applied.

    Throughout our analysis, we assume a spatially flat Universe with adiabatic initial conditions, the latter motivated by the negligible effect of ALP isocurvature perturbations in the mass range considered, see e.g.~\cite{Gorghetto:2021fsn} and the \cref{appendice: initial cond}. Neutrinos are modeled as one massive and two massless species, and the sum of neutrino masses is fixed to the minimal value allowed by flavor oscillation experiments in the normal ordering scenario, $\sum m_\nu = 0.06 \; {\rm eV}$~\cite{deSalas:2020pgw,Esteban:2020cvm,Capozzi:2021fjo}. The neutrino contribution to the effective number of relativistic species is kept fixed at $N_{\rm eff} = 3.044$~\cite{Mangano:2001iu,Bennett:2019ewm,Bennett:2020zkv,Akita:2020szl,Froustey:2020mcq,Drewes:2024wbw}. Any possible ALP contribution to $\dnef$ is automatically taken into account by the Boltzmann code through the integration of the distribution function. The parameter set varied in our analysis is $\{ \omega_b,\, \omega_c,\, 100 \, \theta_s, \, \tau_{\rm reio}, \, \ln(10^{10}A_s), \, n_s, \, \mathcal{N}_\text{ALPs}\}$, where $\omega_b \equiv \Omega_b h^2$ and $\omega_c \equiv \Omega_c h^2$ represent the physical density of baryons and CDM, respectively, $\theta_s$ is the angular size of the acoustic scale at recombination, $\tau_{\rm reio}$ is the reionization optical depth, $A_s$ is the amplitude of the primordial power spectrum at the pivot scale $k_* = 0.05 \; {\rm Mpc}^{-1}$, and $n_s$ is the scalar spectral index. The parameter $\mathcal{N}_{\rm ALPs}$ controls the normalization of the ALP distribution function. For a given ALP mass, we compute the value of $f_a$ (for $q_a =2$) or $A$ (for $q_a=1$) yielding $\Omega_a = 0.26$. The corresponding distribution function defines the reference normalization (for that mass) $\mathcal{N}_{\rm ALPs}=1$. Varying $\mathcal{N}_{\rm ALPs}$ then rescales the distribution function and, consequently, the ALP energy density $\Omega_a$ linearly. For $q_a =2$, this is not exactly equivalent to varying $f_a$, since the latter would also slightly modify the shape of the distribution function through the logarithmic corrections in \cref{eq:distribution_qgr1}, as well as the UV cutoff. We argue however that both effects are small, since the logarithmic corrections are subdominant, and the UV tail of the distribution contributes negligibly to the total energy density. This approach represents a good compromise between faithful and fast modeling.\footnote{We checked that when $f_a$ is varied between the reference value corresponding to $\mathcal{N}_{\rm ALPs}=1$ and 10\% of this value, corresponding to a shift from $\Omega_a \simeq 0.26$ to $\Omega_a \simeq 0.0026$, the error introduced in the energy density by neglecting the variation in the UV cutoff and the one induced by not accounting for the change of shape induced by $\ln_\star$ amount to at most $6.5\%$, across the mass range considered here, with most of the error coming from the latter effect.}

    For the $\Lambda$CDM parameters we adopt uniform priors following the prescriptions of~\cite{SPT-3G:2025bzu}. Instead, for $\mathcal{N}_\text{ALPs}$ we always assume a flat prior ranging from zero up to a value corresponding to $\Omega_a(\mathcal{N}_\text{ALPs})= 1 $. Convergence of the MCMC chains is assessed through the Gelman–Rubin statistic, requiring $R-1 < 0.01$~\cite{Gelman:1992zz}. The posterior distributions are then obtained and analyzed using the \texttt{GetDist} package~\cite{Lewis:2019xzd}. 

    We perform a total of 36 MCMC runs for cosmological models including ALPs from cosmic strings, exploring six different masses, $\log_{10}(m_a/\text{eV}) = \{ -21, -20, -18, -16, -14, -12\}$, and two different tilts, $q_a = \{1,\;  2\}$, distributed as follows. For the tilt $q_a =1$, and each mass, we perform the combinations of datasets (SPA5k+B, SPA3k+B, SPA5k+B+L and SPA3k+B+L), for a total of 24 runs; for  $q_a =2$ and for each mass we only explore the combinations with lensing SPA5k+B+L and SPA3k+B+L, for 12 runs.

    The lower end of the ALP mass range is chosen to ensure that ALPs do not exhibit wave-like behavior on scales probed by CMB experiments. We require the largest de Broglie wavelength in the distribution, corresponding to the lowest momentum considered, to be much smaller than the smallest scale probed by CMB observations. We estimate the latter as $\lambda_{\rm CMB} = 2\pi\chi/\ell_{\rm max} \simeq 5.4 \times 10^{23}$\,m, where $\chi \simeq 4.3 \times 10^{26}$\,m is the comoving distance to recombination and we take $\ell_{\rm max} = 5000$. At the end of the scaling regime, the minimum ALP momentum is $\pmin = (10/3)m_a$, redshifted by a factor $1+z_\star$ as discussed in section~\ref{sec:background}. The maximum de Broglie wavelength today is then
    \begin{equation}
        \lambda_{\rm dB, 0}^{\rm max} = \frac{2\pi(1+z_\star)}{\pmin} \simeq 1.83 \times 10^{21}\, c_z(m_a)\, \left(\frac{m_a}{10^{-21}{\rm\,eV}}\right)^{-1/2}\,{\rm m}\,,
    \end{equation}
    corresponding to the ratio $\lambda_{\rm CMB}/\lambda_{\rm dB, 0}^{\rm max} \simeq 300$ for $m_a=10^{-21}$\,eV. This ratio grows as $m_a^{1/2}$, ensuring that the condition $\lambda_{\rm CMB} \gg \lambda_{\rm dB, 0}$ is satisfied. Additionally, we do not push our analysis to smaller masses due to the fact that very light ALPs should contribute non negligibly to isocurvature fluctuations, as shown in \cref{appendice: initial cond}.

    In our analysis we use the \texttt{CLASS} precision settings adopted in the ACT DR6 analysis of extended cosmological models~\cite{AtacamaCosmologyTelescope:2025nti} in order to precisely model the $C_\ell$ up to $\ell\sim 5000$, with one exception: the parameter $\ell_\text{max}^\text{ncdm}$ at which the Boltzmann hierarchy of non-cold DM is truncated.  We keep it at the \texttt{CLASS} default value of 17, instead of the value of 30 used in the ACT analysis. Being the Boltzmann hierarchy expanded (in the most general case, while the mode is superhorizon) in $N\times \ell_\text{max}^\text{ncdm}$ addends to be evolved separately for every mode, where $N=20$ are the momentum nodes in the quadrature (see \cref{class imp}), this restriction significantly speeds up the computations. To verify that this choice has no impact on the constraints we obtain, we checked that the $\Delta\chi^2$ of the datasets used in our analysis for identical sets of physical parameters, first with $\ell_\text{max}^\text{ncdm}=17$ and then with $\ell_\text{max}^\text{ncdm}=30$ are completely negligible.

\section{Results \label{sec:results}}

    \begin{figure}[t]
        \centering
        \includegraphics[width=1\linewidth]{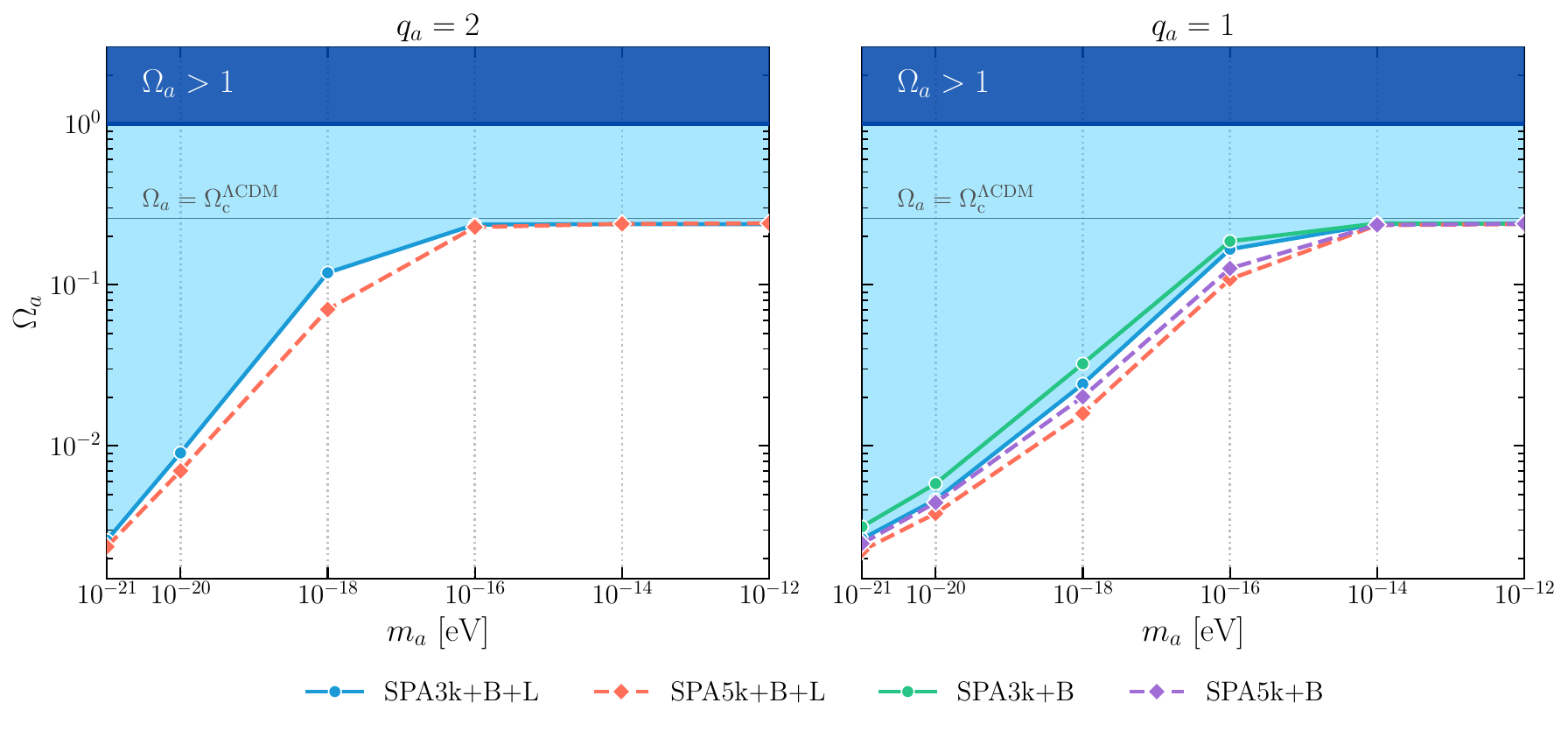}
        \caption{95\% credible intervals for the ALPs density parameter $\Omega_a$ for fixed masses in the range $m_a \in [10^{-21}, 10^{-12}]$ eV, from the dataset combinations described in \cref{sec:methods}. The shaded region shows the region excluded (with 95\% probablity) by SPA3K+B+L. The two panels shows results for $q_a=2$ (left) and $q_a=1$ (right). The circles and diamonds indicate the masses for which we have performed an actual MCMC run. The lines interpolate between these cases.}
        \label{fig: omega alps}
    \end{figure}
    
    \Cref{fig: omega alps} shows a visual overview of the 95\% credible upper bound  for $\Omega_a$ in all the cases we discussed, and is the main result of this work. The bounds are also summarized in \cref{tab: omega a bounds}. The constraints on the DM fraction in ALP, $\Omega_a/\Omega_\mathrm{DM}$ depend in a crucial manner on the ALP mass. For the largest masses in the range considered, the ALP density is essentially bounded by the DM density: $\Omega_a\le\Omega_\mathrm{DM}\simeq{0.26}$. On the other hand, for smaller masses, the constraints become much stronger: at the lower end of the range, we find $\Omega_a< (2\text{--}3)\times 10^{-3}$. This qualitative behaviour, holding for both values of the tilt and for all the dataset combinations considered here, is consistent with the phenomenology discussed in \cref{sec:pheno}: while ALPs behave as cold DM at the high end of the mass range, the warm/hot component of the ALP population, and its distinctive signatures on observables, become increasingly important for lower masses. Comparing the bounds from different dataset combinations we can appreciate how the addition of the smaller angular scales and of the lensing potential data yields stronger constraints for all but the larger masses, thanks to the improved sensitivity to the suppression of matter fluctuations at small scales brought along by the enlarged dataset. In particular, the 95\% upper limit on $\Omega_a$ shrinks by as much as $30\%$ (at $m_a=10^{-18}\,\mathrm{eV}$, $q_a=1$) when adding lensing from {\it Planck}, and by as much as $43\%$ (at $m_a=10^{-18}\,\mathrm{eV}$, $q_a=2$) when increasing $\ell_\mathrm{max}$ to 5000. We have verified that, throughout the parameter region allowed within 95\% credible intervals, the contribution of ALPs to $\Delta N_{\rm eff}$ is negligible (at most of order $10^{-2}$) over all our mass spectrum. Therefore, the constraints derived in our analysis are not driven by the additional radiation density associated with the ALP population.

    \begin{table}
        \centering
        \small
        \renewcommand{\arraystretch}{1.4}
        95\% credible upper bound on $\Omega_a$ for $q_a = 2$ \\
        \vskip 1mm
        \begin{tabular} { l c c c c c c}
            \hline\hline
            & {\boldmath$10^{-12}\,\mathrm{eV}$} & {\boldmath$10^{-14}\,\mathrm{eV}$} & {\boldmath$10^{-16}\,\mathrm{eV}$} & {\boldmath$10^{-18}\,\mathrm{eV}$} & {\boldmath$10^{-20}\,\mathrm{eV}$} & {\boldmath$10^{-21}\,\mathrm{eV}$}\\
            \hline
            {\bf SPA3k+B+L} & $< 0.24$ & $< 0.24$ & $< 0.24$ & $< 0.12$ & $< 0.0090$ & $< 0.0026$\\
            {\bf SPA5k+B+L} & $< 0.24$ & $< 0.24$ & $< 0.23$ & $< 0.070$ & $< 0.0070$ & $< 0.0024$\\
            \hline\hline
        \end{tabular} \\
        \vspace{0.5cm}
        95\% credible upper bound on $\Omega_a$ for $q_a = 1$ \\
        \vskip 1mm
        \begin{tabular} { l c c c c c c}
            \hline\hline
            & {\boldmath$10^{-12}\,\mathrm{eV}$} & {\boldmath$10^{-14}\,\mathrm{eV}$} & {\boldmath$10^{-16}\,\mathrm{eV}$} & {\boldmath$10^{-18}\,\mathrm{eV}$} & {\boldmath$10^{-20}\,\mathrm{eV}$} & {\boldmath$10^{-21}\,\mathrm{eV}$}\\
            \hline
            {\bf SPA3k+B+L} & $< 0.24$ & $< 0.24$ & $< 0.17$ & $< 0.024$ & $< 0.0046$ & $< 0.0026$\\
            {\bf SPA5k+B+L} & $< 0.24$ & $< 0.23$ & $< 0.11$ & $< 0.016$ & $< 0.0038$ & $< 0.0022$\\
            {\bf SPA3k+B} & $< 0.24$ & $< 0.24$ & $< 0.19$ & $< 0.032$ & $< 0.0058$ & $< 0.0031$\\
            {\bf SPA5k+B} & $< 0.24$ & $< 0.24$ & $< 0.13$ & $< 0.020$ & $< 0.0044$ & $< 0.0025$\\
            \hline\hline
        \end{tabular} \\
        \caption{95\% credible upper bounds on $\Omega_a$ for $q_a=2$ (upper panel) and $q_a=1$ (lower panel), obtained using different combinations of data sets. The bounds are reported for different ALP masses.}
        \label{tab: omega a bounds}
    \end{table}

\begin{figure}
        \centering
        \includegraphics[width= \linewidth]{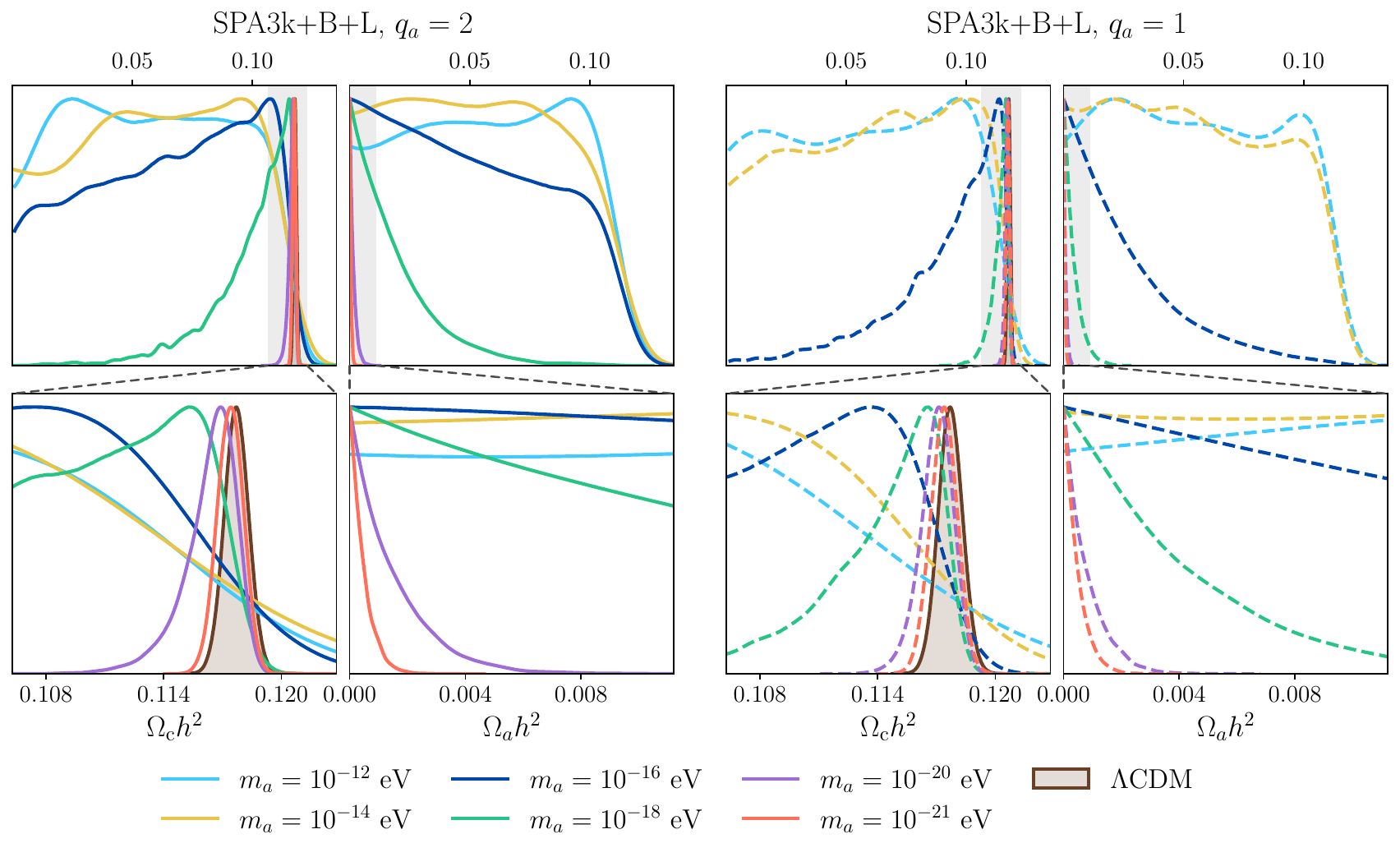}
        \caption{Comparison of posteriors for $q_a=1$ and $q_a=2$, one of the most constraining dataset. We plot $\Lambda$CDM in brown and filled for better visibility in both cases. \textit{Left panel:} plots corresponding to $q_a=2$, solid lines correspond to SPA3k+B+L. Bottom figures are a blowup of the region interesting for the lightest masses. \textit{Right panel:} same as the left panel for $q_a=2$.}
        \label{fig:triangle_comparison}
    \end{figure}

    Comparing the two panels of \cref{fig: omega alps}, or the values in \cref{tab: omega a bounds}, allows to assess how the constraints depend on the spectral tilt. The scale-invariant ($q_a=1$) case is more constrained (up to a factor 5) than the $q=2$ case for masses between $10^{-20}$ and $10^{-16}$\,eV. We trace this to the fact the spectrum is IR dominated in the latter case, so that, all other parameters -- mass, abundance -- being fixed, we expect the scale-invariant distribution to be hotter than the one with $q=2$. We expect this to be true in general, i.e. that smaller tilts (including UV-dominated spectra with $q_a<1$, currently disfavored by lattice simulations) would be more constrained. We take however the fact that the bounds are  relatively robust between the two tilts considered here as an indication that the magnitude of the bound for $1\le q_a\le 2$ only secondarily depends on the details of the distribution function, and rather is governed by general characteristics, most importantly the momentum support, the quasi-scale invariance and the total amount of redshift the species undergoes from the time the network is destroyed, and of course on the mass. We explicitly show the one dimensional posteriors of $\Omega_a h^2$ and $\Omega_c h^2$ in \cref{fig:triangle_comparison}. The posteriors show that an increase in the ALP abundance is compensated by a decrease in the CDM abundance. Moreover, the amount of ALPs as DM that can be tolerated increases proportionally to the increase in the mass, as expected in light of the discussion on the \vg coldness'' of more massive ALPs, and thus their increased similarity with CDM.

\begin{figure}[t]
        \centering
        \includegraphics[width=0.8\linewidth]{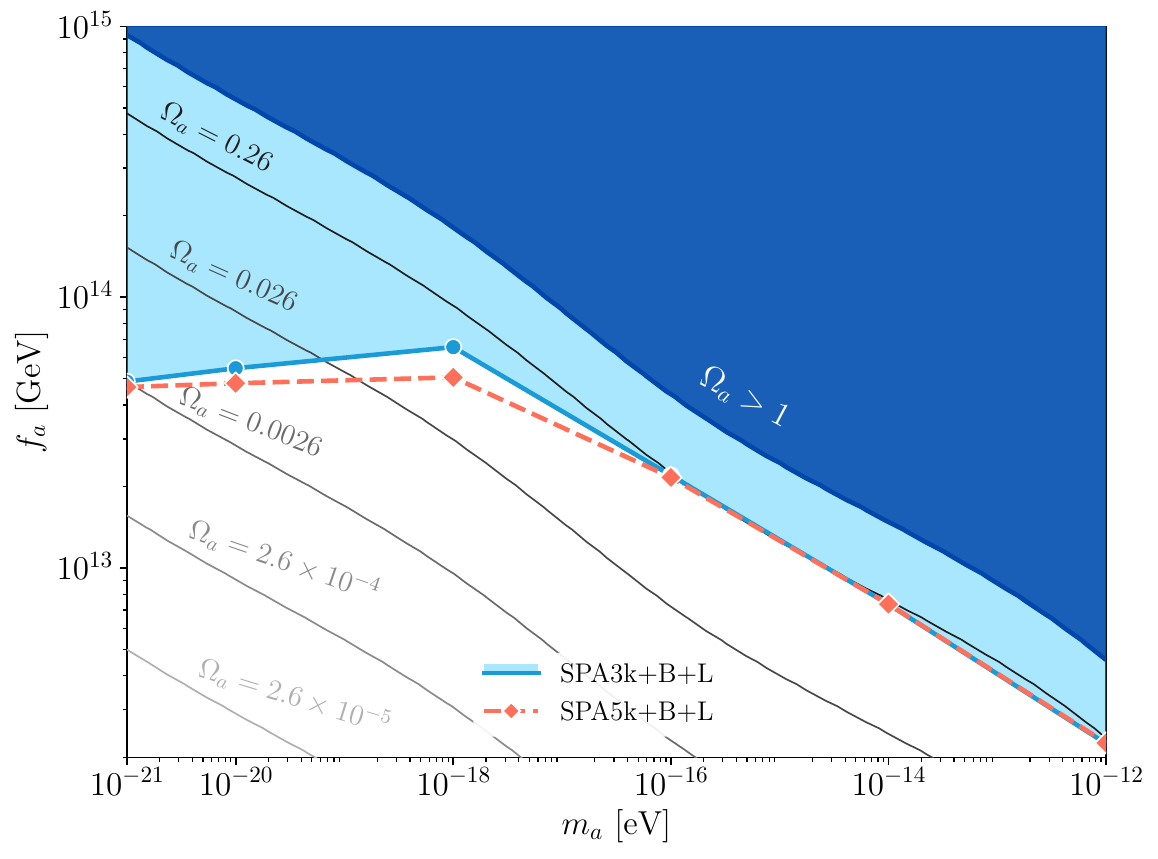}
        \label{fig:mafa_cmb}
        \caption{Constraints on $m_a$-$f_a$ parameter space, for the case $q_a = 2$. For a comparison we refer to~\cite{Gorghetto:2025uls}. The gray lines represent $\Omega_a$ computed numerically, almost (but not exactly) proportional to $\Omega_a \propto f_a^2 \sqrt{m_a}$. Orange and light blue bounds, very close to each other, are either from SPA3k+B+L (baseline) or SPA5k+B+L, while the white region is the one currently allowed. Bounds are derived computing $\Omega_a$ numerically, taking in consideration all effects of $f_a$ and of the changing $g_*, g_{*s}$. }
        \label{fig:mafa_combined}
    \end{figure}

    For the case $q_a =2$, the bounds on $\Omega_a$ can be translated into constraints in the $f_a $--$\,m_a$ parameter space. In \cref{fig:mafa_combined}, fixing $q_a =2$, we numerically compute $\Omega_a$ by integrating the distribution function over a grid of $f_a$ and $m_a$ values. Alternatively, an approximate estimate of the ALP abundance as a function of $m_a$ and $f_a$ can be obtained by exploiting the scaling $\Omega_a \propto f_a^2 \sqrt{m_a}$. This gives
    \begin{equation} \label{omega bounds}
        \Omega_a \sim 0.5 \left(\frac{f_a}{10^{14} \, \text{GeV}}\right)^2\left(\frac{m_a}{10^{-18} \,\text{eV}}\right)^{1/2}.
    \end{equation}
    We emphasize that this approximate relation holds under the assumptions of our modeling, in particular upon the choice of $q_a$.

\section{Discussion and conclusions} \label{sec:conclusions}

    We have studied the cosmological phenomenology of ALPs produced by a network of cosmic strings, and derived constraints on their abundance within the mass range $10^{-21}$--$10^{-12}$\,eV, using state-of-the art measurements of CMB temperature and polarization anisotropies, CMB lensing, and BAO. 

    Given the uncertainty from lattice simulations of cosmic string networks in the spectral tilt of the instantaneous emission rate, $q_a$, we have explored both a scale-invariant spectrum with $q_a=1$ and an IR-dominated spectrum with $q_a=2$. For both benchmarks, we find that the constraints depend strongly on the ALP mass: while ALPs can account for all of the DM at the upper end of the mass range, their contribution is limited at smaller masses. For $m_a = 10^{-18}$\,eV, our baseline dataset bounds the ALP fraction in DM to less than 9\% for $q_a=1$ and 47\% for $q_a=2$, at 95\% credibility. For $m_a = 10^{-21}$\,eV, the limits are below 1\%, for both spectral tilts. This behavior is related to the large span in momentum of the distribution function, and the mass dependence of the network disruption epoch. For smaller ALP masses, the later destruction of the string network leaves less time for the emitted particles to redshift, resulting in a mixture of cold and warm/hot DM.

    We find systematically tighter constraints for $q_a =1$, as expected from the greater population of high-momentum modes, which enhances the suppression from free-streaming in the matter power spectrum. In any case, the qualitative behavior of the constraints is compatible between the two benchmark scenarios. The largest difference is observed for $m_a = 10^{-18}$\,eV, where the constraints for the scale-invariant case are a factor $\sim 5$ stronger than those for the IR-dominated spectrum.

    For the $q_a=2$ benchmark, we translate the abundance constraints into upper limits on the ALP decay constant $f_a$. At 95\% credibility, we obtain $f_a \lesssim 2\times 10^{12}$\,GeV for $m_a = 10^{-12}$\,eV, and $f_a \lesssim 5\times 10^{13}$\,GeV for $m_a = 10^{-21}$\,eV. For masses $m_a \lesssim 10^{-16}$\,eV, these bounds are stronger than those obtained by imposing only the relic abundance condition, $\Omega_a \le \Omega_{\rm DM}$. Neglecting the effects of ALP free-streaming can lead to a severe overestimate of the allowed parameter space. For example, given the approximate scaling $\Omega_a\propto f_a^2$, applying the naive constraint $\Omega_a \le \Omega_\mathrm{DM}$ for $m_a = 10^{-21}$\,eV yields a bound on $f_a$ approximately one order of magnitude weaker than the one obtained here. 

    Our analysis extends previous studies of CMB constraints that treat ALPs as cold DM~\cite{Gorghetto:2021fsn}, and complements the recent results in~\cite{Gorghetto:2025uls}, which considers ALP isocurvature perturbations using UV luminosity functions, Lyman-$\alpha$ observables and CMB data over the mass range $10^{-30}$--$10^{-18}$\,eV. While our constraints are mainly driven by free-streaming, we obtain competitive bounds in the overlapping mass range, using observables that are relatively insensitive to nonlinear evolution.

    Our analysis neglects nonlinear effects in the matter power spectrum on the lensed CMB anisotropies and lensing potential, given the absence of a reliable recipe to compute such corrections. To assess the sensitivity of our constraints, we have compared several dataset combinations. The most conservative choice, SPA3k+B, uses only current observations of the lensed CMB at scales $\ell\le 3000$ that are expected to be only weakly affected by nonlinearities. We have also considered combinations extending the multipole range to $\ell \le 5000$, including the lensing potential likelihood, or both.  For $q_a=1$, the upper limits obtained with SPA3k+B are at most a factor 2.5 weaker than those from the less conservative combinations, with the largest difference achieved again for $m_a = 10^{-18}$\,eV.  We also find that including \textit{Planck} lensing reconstruction data only mildly improves the constraints, which are thus driven by the indirect effect of lensing on temperature and polarization anisotropies. For $q_a=2$, we did not perform an equivalent comparison including the SPA3k+B dataset without lensing, although we expect similar results given that the non-cold population is less relevant for an IR-dominated distribution and is the component mainly constrained by small-scale and lensing observables when all other parameters are held fixed. A proper treatment of nonlinearities is however required to robustly extract the full information encoded in CMB observables, and is left for future developments of this work.

    Given that the bounds on the model are driven by the peculiar evolution of matter perturbations sourced by ALPs from cosmic strings, we expect that including also more sensitive probes of the large-scale structure, like galaxy clustering and galaxy weak lensing observations, together with their cross-correlation with CMB observables, would significantly improve the limits reported here. Also in the case of galaxy observables, a proper treatment of nonlinearities would greatly improve the constraining power for the model considered here.

    Beyond the bounds we have derived, this work also illustrates the role of the DM distribution function. Cosmological observables like CMB lensing can be sensitive to the presence of a species with a sufficiently nonstandard distribution function, for example one with a significant free-streaming component. Our analysis, including the numerical pipeline developed to model the distribution function of ALPs, may also serve as a guide for modeling other nonthermal species.

\acknowledgments
RI thanks M.\ Hindmarsh, J.\ Correia and A.\ Drew for useful discussions.
We acknowledge the use of computing facilities provided by the INFN theory group (I.S. InDark) at CINECA.
NB is supported by the Spanish grants PID2023-147306NB-I00 and CEX2023-001292-S (MCIU/AEI/10.13039/501100011033).
LC acknowledges the financial support provided through national funds by FCT -- Funda\c{c}\~ao para a Ci\^encia e Tecnologia, I.P., with DOI identifiers 10.54499/2023.11681.PEX, 10.54499/UID/04564/\allowbreak 2025 and by the project 2024.00249.CERN funded by measure RE-C06-i06.m02 -- ``Reinforcement of funding for International Partnerships in Science, Technology and Innovation'' of the Recovery and Resilience Plan -- RRP, within the framework of the financing contract signed between the Recover Portugal Mission Structure (EMRP) and the Foundation for Science and Technology I.P.\ (FCT), as an intermediate beneficiary. 
NB and ML acknowledge funding by the European Union (ERC, RELiCS, project number 101116027). Views and opinions expressed are however those of the authors only and do not necessarily reflect those of the European Union or the European Research Council Executive Agency. Neither the European Union nor the granting authority can be held responsible for them.
LV acknowledges support from the Istituto Nazionale di Fisica Nucleare (INFN) through the Commissione Scientifica Nazionale 4 (CSN4), Iniziativa Specifica ``Quantum Universe'' (QGSKY), and the Commissione Scientifica Nazionale 2 (CSN2), experiment QUAX. We acknowledge support by Commissione Scientifica Nazionale 4 (CSN4) and the Research Network InDark funded by Istituto Nazionale di Fisica Nucleare (INFN), and by the ASI (Italian Space Agency) Grant 2020-9-HH.0 (participation in LiteBIRD, phase A).

\appendix

\section{Details of the \texttt{CLASS} implementation}\label{class imp}

In this pedagogical Appendix, we describe the theoretical framework and numerical choices underlying our implementation of ALP momentum distributions in \class\ \cite{Lesgourgues:2011rh}. We first explain how the distribution function enters the background equations and the integro-differential equations governing linear perturbations. We then discuss numerical quadrature schemes for the momentum integrals and their implementation in Boltzmann codes such as \class. Finally, we present our implementation choices and validation tests.

\subsection{The phase-space distribution function of non-cold relics}

The phase-space distribution function $\mathcal{F}$ of a non-cold relic determines its contribution to the background energy-momentum tensor and its perturbations~\cite{Ma:1995ey}. Its evolution is governed by the Boltzmann equation. In the following, we consider a species of constant mass $m$ undergoing collisionless evolution.

We write the distribution function in terms of comoving position $\mathbf{x}$, comoving momentum $\mathbf{q}=a\mathbf{p}$, and conformal time $\tau$, where $\mathbf{p}$ is the physical momentum. Splitting it into a homogeneous background and a linear perturbation gives
\begin{equation}
    \mathcal{F}(\mathbf{x},\mathbf{q},\tau)
    = \mathcal{F}_0(q)
      \left[1+\Psi(\mathbf{x},\mathbf{q},\tau)\right].
\end{equation}
Background homogeneity and isotropy imply that $\mathcal{F}_0$ depends on the phase-space coordinates only through $q=|\mathbf{q}|$. Collisionless evolution further implies that $\mathcal{F}_0(q)$ is independent of time. 

We absorb internal-state degeneracies and phase-space normalization factors into $\mathcal{F}$, so that the background number density is $n_\mathrm{ncdm}=a^{-3}\int {\rm d}^3q\,\mathcal{F}_0(q)$. Here the subscript $\mathrm{ncdm}$ stands for non-cold dark matter. Defining the comoving energy
\begin{equation} \label{eq:comoving_energy}
    \epsilon(q,a)=aE=\sqrt{q^2+a^2m^2},
\end{equation}
the background energy density and pressure are
\begin{align}
    \label{background quantities}
    \rho_\mathrm{ncdm}(a)
    &= 4\pi a^{-4}\int_0^\infty {\rm d}q\,
       q^2\epsilon(q,a)\mathcal{F}_0(q), \\
    P_\mathrm{ncdm}(a)
    &= \frac{4\pi}{3}a^{-4}\int_0^\infty {\rm d}q\,
       \frac{q^4}{\epsilon(q,a)}\mathcal{F}_0(q).
\end{align}
The angular integrals have been performed using background isotropy. Once $m$ and $\mathcal{F}_0(q)$ are specified, these expressions determine the energy density and pressure at any scale factor $a$, without requiring a separate evolution equation for the background distribution. Together with the contributions from the other species, $\rho_\mathrm{ncdm}(a)$ enters the first Friedmann equation, which determines the evolution of the scale factor.

To describe scalar perturbations, we Fourier-transform the spatial dependence of $\Psi$ and expand its momentum-direction dependence in Legendre polynomials:
\begin{equation}
    \Psi(\mathbf{k},\mathbf{q},\tau)
    = \sum_{\ell=0}^{\infty}
      (-i)^\ell(2\ell+1)\Psi_\ell(\mathbf{k},q,\tau)
      P_\ell(\hat{\mathbf{k}}\cdot\hat{\mathbf{q}}).
\end{equation}
For scalar perturbations with axially symmetric initial conditions, the dependence on $\hat{\mathbf{q}}$ enters only through $\hat{\mathbf{k}}\cdot\hat{\mathbf{q}}$. We work in the conformal Newtonian gauge, with metric
\begin{equation}
    {\rm d}s^2
    = a^2(\tau)\left[
      -(1+2\psi){\rm d}\tau^2
      +(1-2\phi){\rm d}\mathbf{x}^2
      \right].
\end{equation}
The collisionless Boltzmann equation then yields the infinite hierarchy
\begin{subequations}
\label{boltzmann hierarchy}
\begin{align}
    \dot{\Psi}_0
    &= -\frac{qk}{\epsilon}\Psi_1
       -\dot{\phi}\frac{{\rm d}\ln\mathcal{F}_0}{{\rm d}\ln q}, \\
    \dot{\Psi}_1
    &= \frac{qk}{3\epsilon}(\Psi_0-2\Psi_2)
       -\frac{\epsilon k}{3q}\psi
        \frac{{\rm d}\ln\mathcal{F}_0}{{\rm d}\ln q}, \\
    \dot{\Psi}_\ell
    &= \frac{qk}{(2\ell+1)\epsilon}
       \left[\ell\Psi_{\ell-1}-(\ell+1)\Psi_{\ell+1}\right],
       \qquad \ell\geq 2.
\end{align}
\end{subequations}
Here $k=|\mathbf{k}|$, overdots denote derivatives with respect to conformal time, and $\epsilon=\epsilon(q,a)$. We suppress the arguments $(\mathbf{k},q,\tau)$ of $\Psi_\ell$ and $(\mathbf{k},\tau)$ of the metric potentials.
In numerical implementations, the hierarchy is truncated at a finite
multipole $\ell_\mathrm{max}^\mathrm{ncdm}$ and supplemented by a closure relation. The closure relation suggested in Ref.~\cite{Ma:1995ey} and implemented in \texttt{CLASS} reads, for vanishing spatial curvature, $\Psi_{\ell+1}=(2\ell+1)\epsilon/\left(qk\tau \right)\Psi_\ell - \Psi_{\ell-1}$. The evolution equations are solved at a discrete set of $n_q$ comoving momenta $q_i$. Therefore, each ncdm species contributes $n_q (\ell_\mathrm{max}^\mathrm{ncdm}+1)$ equations to the coupled Einstein-Boltzmann system. 

The potentials $\phi$ and $\psi$ obey the linearized Einstein equations, which are sourced by perturbations to the total energy-momentum tensor. The non-cold relic contributes through its energy-density and pressure perturbations, velocity divergence $\theta_\mathrm{ncdm}$, and scalar anisotropic stress $\sigma_\mathrm{ncdm}$. With the above Legendre convention, these quantities are
\begin{subequations}
\label{perturbation quantities}
\begin{align}
    \delta\rho_\mathrm{ncdm}
    &= 4\pi a^{-4}\int_0^\infty {\rm d}q\,
       q^2\epsilon\,\mathcal{F}_0(q)\Psi_0, \\
    \delta P_\mathrm{ncdm}
    &= \frac{4\pi}{3}a^{-4}\int_0^\infty {\rm d}q\,
       \frac{q^4}{\epsilon}\mathcal{F}_0(q)\Psi_0, \\
    \theta_\mathrm{ncdm}
    &= \frac{4\pi a^{-4}k}
       {\rho_\mathrm{ncdm}+P_\mathrm{ncdm}}
       \int_0^\infty {\rm d}q\,
       q^3\mathcal{F}_0(q)\Psi_1, \\
    \sigma_\mathrm{ncdm}
    &= \frac{8\pi a^{-4}}
       {3(\rho_\mathrm{ncdm}+P_\mathrm{ncdm})}
       \int_0^\infty {\rm d}q\,
       \frac{q^4}{\epsilon}\mathcal{F}_0(q)\Psi_2.
\end{align}
\end{subequations}

These momentum integrals source the evolution of the metric potentials, which in turn enter the evolution equations for the distribution-function multipoles. Consequently, the hierarchy in \cref{boltzmann hierarchy}, the linearized Einstein equations, and the perturbation equations for the other species form a coupled system of integro-differential equations. The momentum integrals in \cref{perturbation quantities} must therefore be evaluated as part of the perturbation evolution. In practice, these integrals are evaluated by numerical quadrature, using the multipoles $\Psi_\ell$ computed at nodes given by the sampled momenta $q_i$.

We finally note that, for modes well inside the horizon, \class\ can switch to an approximate
fluid description, evolving only the density contrast, velocity divergence,
and anisotropic stress of each ncdm species.

\subsection{Integration of the distribution function}

    \class\  solves the background evolution and the perturbations evolution in two dedicated modules, \texttt{background.c} and \texttt{perturbations.c}. Both modules need to perform integrals over the distribution function, see \cref{background quantities,perturbation quantities}. Note however that the background and perturbation calculations have different requirements and are subject to different constraints.
    The background calculation needs to be very precise; however, the code spends a small fraction of the computing time in \texttt{background.c}. Therefore the background quadrature strategy can be numerically expensive for a single integral evaluation, without sacrificing the speed of the code. For what concerns the perturbations, instead, the distribution function has to be evolved for each momentum node $q_i$, therefore the number of nodes $n_q$ impacts heavily on the speed of the code. In the perturbation module it is therefore desirable to use a quadrature strategy that allows to reach the desired accuracy using the smallest possible number of nodes.
    
    The publicly available version of the code gives the user some control over the quadrature strategy. By default, \class\ uses an ``automatic quadrature'' mode, where the quadrature strategy is optimized separately for the background and perturbations. Alternative options include Gauss-Laguerre, a trapezoidal rule on $q\in[0,\infty)$, a trapezoidal rule on $q\in[0,\,q_\mathrm{max})$ (with user-specified $q_\mathrm{max}$). In this case, however, the strategy, including the sampling, is shared by both the background and the perturbations. The settings for the quadrature strategy can be specified independently for each ncdm species. 

    Two ncdm species are present in the models considered in this work: neutrinos and ALPs from string networks. For neutrinos, we use the Gauss-Laguerre quadrature already implemented in \class. For ALPs, we found that none of the quadrature strategies implemented in \class\ was fit for our purposes, given the highly non-thermal shape of the ALP distribution function and the its wide momentum support. In the following section we describe how we address this issue in our modified version of \class.

    \subsection{Implementation of ALP spectra from cosmic strings in \texttt{CLASS}}

        \begin{figure}
            \centering
            \includegraphics[width=\linewidth]{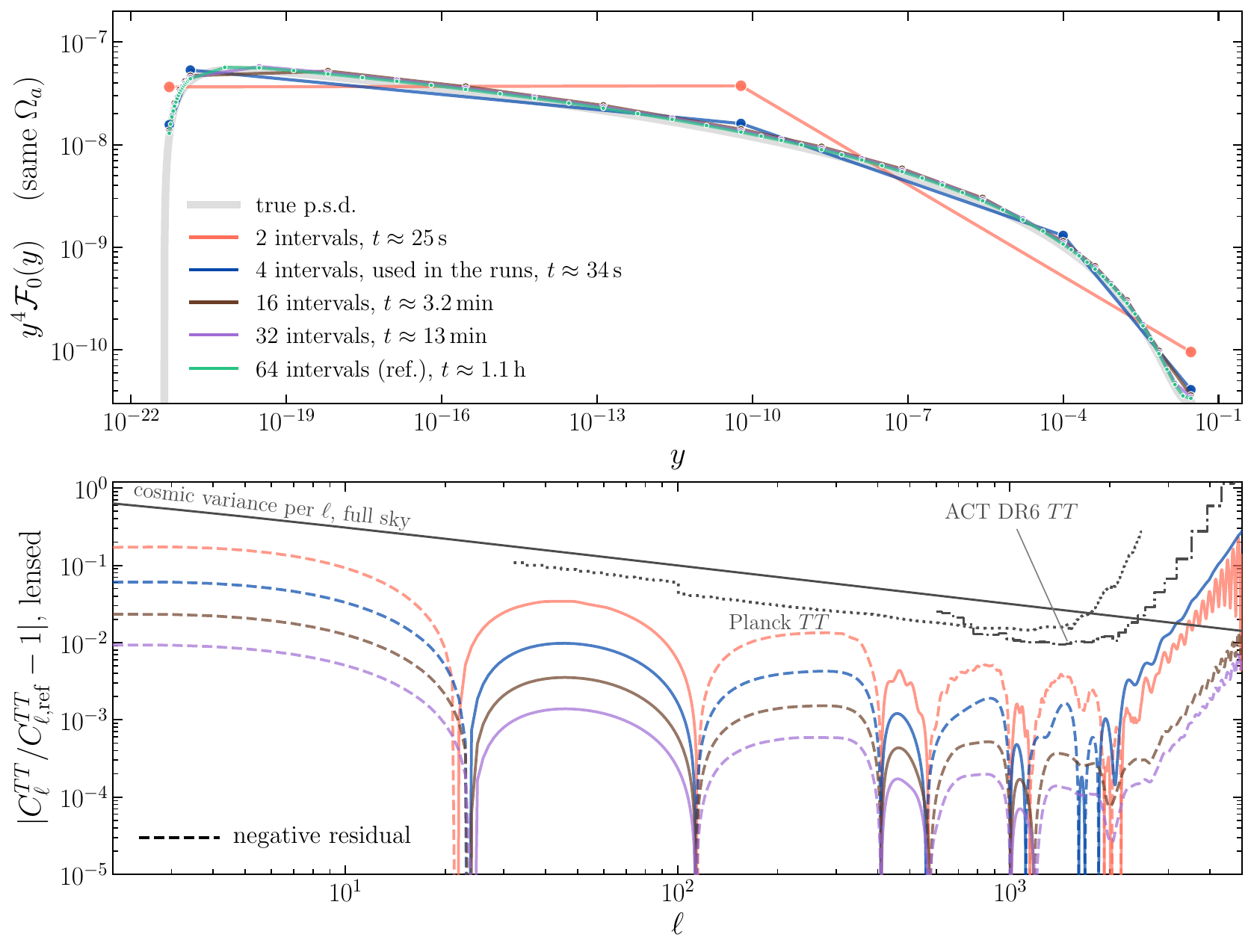}
            \caption{\textit{Top}: Comparison between the exact distribution function (gray solid line) and the broken power-law approximation with $2,\,4,\,16,\,32,\,64$ intervals, in addition to the single interval at the IR cutoff (see text for details). We fix $m_a = 10^{-16} \, \,\mathrm{eV}$ and $q_a = 2$ in this plot. The exact distribution function is computed with a value of $f_a$ corresponding to $\Omega_a = 0.26$. The approximations are rescaled so that they yield the same value of $\Omega_a = 0.26$. The legend reports execution times for a single \class call on a 16 CPUs core, using 4 quadrature nodes in each interval. \textit{Bottom} Relative difference between lensed CMB  temperature power spectra corresponding to the distribution functions in the top panel, using the 64-intervals case (260 nodes) as a reference. We include massive neutrinos and fix all non-ALP cosmological parameters at their fiducial values. Dashed lines signal where the difference is negative. The solid grey line is the cosmic variance per multipole, $\sqrt{2/(2\ell+1)}$; the dotted and dash-dotted steps are the relative $TT$ bandpower errors of {\it Planck} 2018 (\texttt{plik-lite}) and ACT DR6, respectively.}
            \label{fig: interpolation snapshot}
        \end{figure} 

        In the following we describe our implementation of ALPs with a highly nonthermal distribution functions in \class. We will refer to $y = a_\star q/T_{\gamma,0}$ as the dimensionless momentum variable of \texttt{CLASS}, where $T_{\gamma,0} \simeq 2.35 \times 10^{-4}$\,eV is the present CMB temperature.

        First, given the ALP mass $m_a$ and tilt $q_a$ of the instantaneous emission rate, we compute the value of $f_a$ (for $q_a>1$) or $A$ (for $q_a=1$) (see \cref{eq:distribution_qgr1} and \cref{eq:distribution_qeq1}) that yields $\Omega_a = 0.26$ . This completely fixes the background distribution function to a reference that we denote $\mathcal{F}_0^{\rm ref}(y)$, with $y\in\left[ y_\text{IR},  y_\text{UV} \right]$, where  $y_\text{IR}$ and $y_\text{UV}$ are the $y$ cutoffs corresponding to $p_\mathrm{min}^\star$ and $p_\mathrm{max}^\star$. Then, we approximate the distribution function with a piecewise continuous function, in particular, a broken power law. A distinct quadrature strategy is then used for each individual piece. This is dictated by the necessity to keep some control over the position of the quadrature nodes, in particular to ensure that, during the regime where the warm component of ALPs is relevant, the region of momenta $q\sim\tilde a m_a$ is always populated by quadrature nodes. This broken power-law approximation is finally hard-coded in \class.

        We choose the number of power-laws, and their edges, on the base of a trade-off between accuracy and speed of the code. In the upper panel of \cref{fig: interpolation snapshot}, we show, for a representative case the distribution function for different choices of the number of power laws, from 2 to 64, and the corresponding execution times in \class. In the lower panel, we show residuals between the corresponding CMB temperature power spectra, taking the 64 intervals case as a reference, and using 4 quadrature nodes in each interval (see below). We find that using 4 intervals provides the best compromise between accuracy and speed.

        In each interval, we use a Gaussian quadrature scheme with weight function $W(y)=\mathcal{F}_0^\text{ref}(y)$ (see below) with $N=4$ nodes. The choice of the weight function is led by the observation that, when the bulk of momenta contributing to the integrals in \cref{perturbation quantities} are either $\gg m_a$ or $\ll m_a$, the comoving energy in \cref{eq:comoving_energy} reduces to $q$ or $a m$, respectively. 
        In these regimes, the integrands therefore reduce to the product of $\mathcal{F}_0^\text{ref}(y)$ times polynomial functions of degree smaller than $2N-1 = 7$, which are integrated exactly within each subdomain by our Gaussian quadrature. 

        The nodes and weights of the quadrature rule in each subdomain are found through the Golub-Welsch algorithm. In particular, we use the implementation of~\cite{FUKUDA2005143}, that accepts an arbitrary weight function and automatically checks the accuracy in the integration.

        Note that, since we perform Monte Carlo runs at fixed mass and spectral tilt, and the only free parameter of the distribution function is its normalization, the quadrature nodes and weights can be pre-computed just once for each parameter estimation run.
  We use the same quadratures for the background as well, after checking that adopting a high precision quadrature strategy with 1000 log-spaced trapezoidal nodes does not affect the spectra.

We conclude this section by discussing an issue related to the fact that the ALP distribution function is not vanishing smoothly at the edges, but instead presents discontinuities in $y_\mathrm{UV}$ and $y_\mathrm{IR}$\footnote{For a related treatment of nonthermal relics, see ref.~\cite{Parichha:2023dqn}, where the evolution equations are implemented in \class, bypassing the numerical integration of the phase-space distribution.}. The issue lies in the derivative term ${\rm d}\ln \mathcal{F}_0/ {\rm d}\ln q$ that appears in both the evolution equations and the initial conditions. For discontinuous distributions, the derivative would yield contributions proportional to a Dirac delta in \cref{boltzmann hierarchy} at the momentum boundaries, which are missed when the derivative is evaluated numerically assuming a smooth distribution. This is potentially problematic for the IR cutoff, but less so for the UV one, because the high-momentum part of the spectrum contributes negligibly to the integrals.  In fact, when the discontinuity is not accounted for, we fail to recover the pure cold DM limit even in the case where the spectrum is completely nonrelativistic at the times relevant for CMB observations. To address this issue, we regularize the infrared cutoff of the ALP distribution with a steep power law, with the momentum integrals evaluated through trapezoidal quadrature on a logarithmic grid.

    \subsection{Gaussian quadratures}

        In a quadrature scheme, a continuous integral is approximated numerically by a discrete sum. For a function $g(x)$ and a weight function $W(x)$, the integral between $\alpha$ and $\omega$ is approximated as
        \begin{equation}
            \int_\alpha^\omega g(x)W(x)\,{\rm d}x \simeq \sum_{i=1}^{N}w_i g(x_i)\,,
        \end{equation}
        where $\{x_i\}$ are the $N$ integration nodes and $\{w_i\}$ the weights. In Gaussian quadrature, the nodes are determined through a set of $N$ orthogonal weighted polynomials $\{P_j\}$ with respect to the weight function $W(x)$, satisfying
        \begin{equation}
            \int_\alpha^\omega W(x)P_i(x)P_j(x)\,{\rm d}x=0\,,\qquad i\neq j\,.
        \end{equation}
        The nodes $\{x_i\}$ are the roots of the polynomial $P_N$ of degree $N$, while the weights are obtained from the corresponding orthogonal polynomials. A Gaussian quadrature with $N$ nodes integrates exactly all polynomials of degree up to $2N-1$, namely
        \begin{equation}
            \int_\alpha^\omega W(x)x^n\,{\rm d}x=\sum_{i=1}^{N}w_i x_i^n,\qquad 0\leq n\leq 2N-1.
        \end{equation}
        Several numerical methods exist to construct Gaussian quadratures for general positive weight functions, including the Golub-Welsch algorithm employed in~\cite{FUKUDA2005143}. One such example is Gauss-Laguerre quadrature, obtained by choosing the weight function $W(x)= e^{-x}$ on the interval $[0,\infty)$. Since thermal distributions, including Fermi-Dirac and Bose-Einstein distributions, approach such a behavior at large values of the momentum, this quadrature is especially suited for modeling thermal relics, including massive neutrinos.

    \subsection{Trapezoidal quadrature}

        Trapezoidal quadratures are built trading the integral for a sum of rect trapezoids which basis are found by evaluating the function at the nodes, and which heights are the differences of contiguous nodes, and are not a type of Gaussian quadrature. They are mostly well behaved for non oscillating and always positive functions, despite being quite expensive numerically. Another key feature of trapezoidal integrals is that in order to reach high accuracy one should choose nodes densely around a region in which the integrand has features, while they can be chosen widely spaced away from features. For an arbitrary set of ordered nodes $\{x_i\} = \{x_0, x_1, ..., x_n\}$, the corresponding trapezoidal weights read 
        \begin{equation} \label{trapezoidal}
            \{w_i\}= \left\{ \frac{x_1-x_0}{2}, \frac{x_2-x_0}{2}, \cdots , \frac{x_{n}-x_{n-2}}{2}, \frac{x_{n}-x_{n-1}}{2}\right\} .
        \end{equation}

\section{Initial conditions}\label{appendice: initial cond}
        
        \begin{figure}
            \centering
            \includegraphics[width=0.8\linewidth]{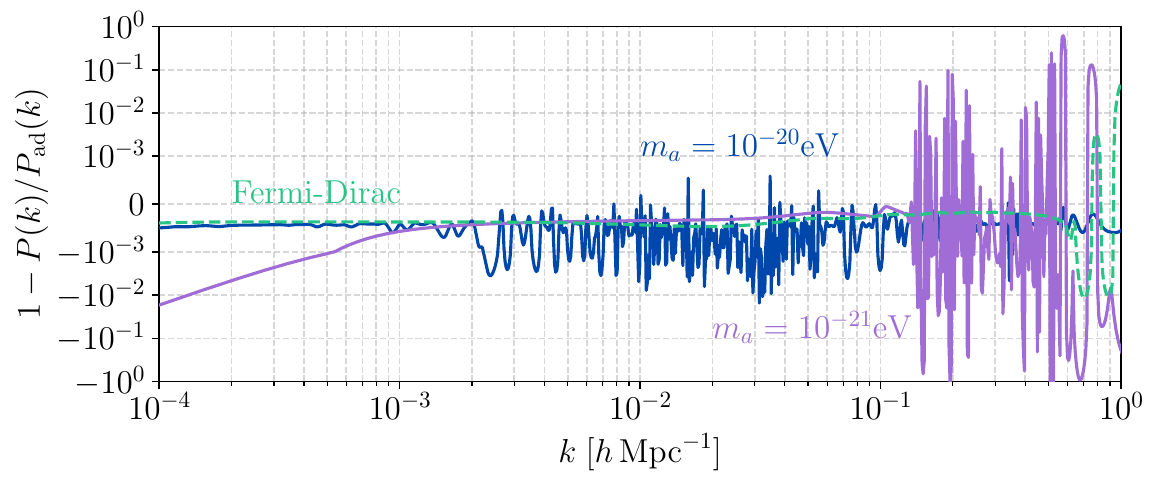}
            \caption{Matter power spectrum for various initial conditions in density and velocity. The bottom panel displays differences with respect to (respectively) the same model but with only adiabatic initial conditions. }
            \label{fig: iso pk}
        \end{figure}
        
        ALPs produced during the scaling regime may carry isocurvature fluctuations, due to the fact that the network need not to be linked to inflaton fluctuations. CMB observations constrain isocurvature fluctuations at the pivot scale $p_{\rm CMB} = 0.05\,{\rm Mpc}^{-1}$, through
        \begin{equation}
            f_{\rm iso}^2 \equiv \frac{\Delta_{\rm iso}^2(p)}{\Delta_{\mathcal{R}}^2(p)}, 
            \qquad 
            \Delta_{\mathcal{R}}^2(p) = A_s \left(\frac{p}{p_{\rm CMB}}\right)^{n_s -1},
        \end{equation}
        with $A_s \simeq 2.2 \times 10^{-9}$.  For $m_a \gtrsim 10^{-28}\,{\rm eV}$, parametrizing $\Delta_a^2(q)=C(q/H_\star)^3$, the fraction of isocurvature perturbations from scaling ALPs can be expressed as
        \begin{equation} \label{isocurvature bound}
            f_{\rm iso} \;\simeq\; 0.2 \left(\frac{\xi_a \log_\star}{3 \times 10^3}\right) \left(\frac{f_a}{5 \times 10^{14}\,{\rm GeV}}\right)^2 \left(\frac{m_a}{10^{-28}\,{\rm eV}}\right)^{-1/4} \left(\frac{C}{2 \times 10^{-5}}\right)^{1/2} \lesssim 0.64,
        \end{equation}
        where the last inequality is the CMB bound for ALPs, obtained in~\cite{Feix_2019} considering a white noise spectrum. Given our mass range and in light of the bounds reported in the main text. $\Omega_\text{DM}$ is more constraining in $f_a$ as the mass grows, while the bound on $N_{\rm eff}$, independent on the mass, is not competitive for this mass range. Finally, $f_{\rm iso}$ is less constraining in $f_a$ as the mass grows. The most stringent bound is obtained setting $m_a \sim 10^{-21}$\,eV, in such case, even pushing $f_a$ much above the $\Omega_{\rm DM}$ bound, we obtain $f_{\rm iso} \ll 0.1$.

        We provide the check that initial conditions can be safely set to adiabatic, in \cref{fig: iso pk}. We confront three Universes in which all of DM is either ALPs from strings (with mass either $m_a = 10^{-21}$ eV or $m_a = 10^{-20}$ eV) and abundance as $\Omega_a=0.26$, or a particle which is Fermi-Dirac distributed and has $\Omega_\text{FD} = 0.26$ ($m_\text{FD} = 0.001$eV, $T_\text{FD} \sim 10^{-5}T_\nu$). We change the initial conditions to be either adiabatic and isocurvature (both temperature and velocity) with a fraction of isocurvature $f_{\rm iso}\sim 0.1$, much larger than the largest value allowed in the bounds reported in~\cite{Gorghetto:2021fsn}. The differences that emerge changing initial conditions and keeping the rest of the physics unchanged is negligibly small on linear scales. We can safely neglect isocurvature initial conditions in our analysis.

\section{Full tables with constraints} \label{appendice: full triangle plots}
    In this Appendix we report the constraints on the cosmological parameters for all the runs. The constraints for $\Lambda$CDM are in \cref{tab: lcdm}, while those for the runs with ALPs from cosmic strings are in \cref{tab: Posteriors for q1 SPA5k+BAO,tab: Posteriors for q1 SPA5k+BAO+Plens,tab: Posteriors for q2 SPA5k+BAO+Plens} for $\ell_\text{max}=5000$ and in \cref{tab: Posteriors for q1 SPA3k+BAO,tab: Posteriors for q1 SPA3k+BAO+Plens,tab: Posteriors for q2 SPA3k+BAO+Plens} for $\ell_\text{max}=3000$.

\begin{table}[h]
    \renewcommand{\arraystretch}{1.5}
    \centering
    {\normalsize \textbf{\boldmath $\Lambda$CDM}\par}
    \vspace{3pt}
    \resizebox{0.9\textwidth}{!}{%
    \begin{tabular}{ l c c c c }
        \hline
        \hline
            & \textbf{SPA5k+B}
            & \textbf{SPA5k+B+L}
            & \textbf{SPA3k+B}
            & \textbf{SPA3k+B+L} \\
        \hline
            {\boldmath$\Omega_\mathrm{b}h^2$} & 
            $0.022503 \pm 0.000094$ & 
            $0.022505 \pm 0.000092$ & 
            $0.022493 \pm 0.000092$ & 
            $0.022493 \pm 0.000097$ \\

            {\boldmath$\Omega_\mathrm{c}h^2$} & 
            $0.11757 \pm 0.00063$ & 
            $0.11759 \pm 0.00062$ & 
            $0.11764 \pm 0.00064$ & 
            $0.11771 \pm 0.00063$ \\

            {\boldmath$100\theta_\mathrm{s}$} & 
            $1.04177 \pm 0.00022$ & 
            $1.04177 \pm 0.00022$ & 
            $1.04179 \pm 0.00022$ & 
            $1.04179 \pm 0.00022$ \\

            {\boldmath$\ln(10^{10} A_\mathrm{s})$} & 
            $3.061^{+0.012}_{-0.014}$ & 
            $3.061 \pm 0.012$ & 
            $3.063^{+0.013}_{-0.014}$ & 
            $3.064 \pm 0.012$ \\

            {\boldmath$n_\mathrm{s}$} & 
            $0.9743 \pm 0.0030$ & 
            $0.9743 \pm 0.0029$ & 
            $0.9733 \pm 0.0030$ & 
            $0.9731 \pm 0.0030$ \\

            {\boldmath$\tau_\mathrm{reio}$} & 
            $0.0651^{+0.0060}_{-0.0068}$ & 
            $0.0650^{+0.0057}_{-0.0063}$ & 
            $0.0657^{+0.0062}_{-0.0071}$ & 
            $0.0663 \pm 0.0062$ \\
        \hline
        \hline
    \end{tabular}
    }
    \caption{Constraints on $\Lambda$CDM parameters from SPA+B and SPA+B+L datasets with $\ell_\text{max}=5000$ (SPA5k) and $\ell_\text{max}=3000$ (SPA3k). Bounds are 68\% credible intervals for detected parameters and 95\% upper limits otherwise.}
    \label{tab: lcdm}
\end{table}

\begin{table}[h]
    \renewcommand{\arraystretch}{1.5}
    \centering
    {\small \textbf{\boldmath $q_a = 1$, \; SPA5k+B}\par}
    \vspace{3pt}
    \resizebox{\textwidth}{!}{%
    \begin{tabular}{ l c c c c c c }
        \hline
        \hline
            & {\boldmath $10^{-12} \, \mathrm{eV}$}
            & {\boldmath $10^{-14} \, \mathrm{eV}$}
            & {\boldmath $10^{-16} \, \mathrm{eV}$}
            & {\boldmath $10^{-18} \, \mathrm{eV}$}
            & {\boldmath $10^{-20} \, \mathrm{eV}$}
            & {\boldmath $10^{-21} \, \mathrm{eV}$} \\
        \hline
            {\boldmath$\Omega_\mathrm{b}h^2$} & 
            $0.022532 \pm 0.000094$ & 
            $0.022530 \pm 0.000095$ & 
            $0.022509 \pm 0.000091$ & 
            $0.022502 \pm 0.000092$ & 
            $0.022502 \pm 0.000091$ & 
            $0.022499 \pm 0.000095$ \\

            {\boldmath$\Omega_\mathrm{c}h^2$} & 
            $< 0.112$ & 
            $0.066^{+0.049}_{-0.028}$ & 
            $0.0977^{+0.020}_{-0.0042}$ & 
            $0.1145^{+0.0033}_{-0.00097}$ & 
            $0.11699^{+0.00098}_{-0.00072}$ & 
            $0.11733 \pm 0.00071$ \\

            {\boldmath$100\theta_\mathrm{s}$} & 
            $1.04181 \pm 0.00022$ & 
            $1.04180 \pm 0.00023$ & 
            $1.04179 \pm 0.00022$ & 
            $1.04177 \pm 0.00023$ & 
            $1.04179 \pm 0.00022$ & 
            $1.04176 \pm 0.00023$ \\

            {\boldmath$\ln(10^{10} A_\mathrm{s})$} & 
            $3.065^{+0.013}_{-0.014}$ & 
            $3.065^{+0.012}_{-0.014}$ & 
            $3.066^{+0.013}_{-0.015}$ & 
            $3.065^{+0.012}_{-0.015}$ & 
            $3.066^{+0.012}_{-0.014}$ & 
            $3.065^{+0.012}_{-0.014}$ \\

            {\boldmath$n_\mathrm{s}$} & 
            $0.9739 \pm 0.0029$ & 
            $0.9739 \pm 0.0030$ & 
            $0.9737 \pm 0.0029$ & 
            $0.9733 \pm 0.0030$ & 
            $0.9732 \pm 0.0030$ & 
            $0.9735 \pm 0.0030$ \\

            {\boldmath$\tau_\mathrm{reio}$} & 
            $0.0664^{+0.0061}_{-0.0072}$ & 
            $0.0666^{+0.0060}_{-0.0070}$ & 
            $0.0673^{+0.0063}_{-0.0073}$ & 
            $0.0671^{+0.0060}_{-0.0075}$ & 
            $0.0672^{+0.0061}_{-0.0072}$ & 
            $0.0669^{+0.0061}_{-0.0072}$ \\[3pt]
        \hdashline
            {\boldmath$\Omega_{a}h^2$} & 
            $< 0.110$ & 
            $< 0.108$ & 
            $< 0.0581$ & 
            $< 0.00927$ & 
            $< 0.00204$ & 
            $< 0.00114$ \\
        \hline
        \hline
    \end{tabular}
    }
    \caption{Constraints on the cosmological parameters and on $\Omega_a h^2$ for ALPs from cosmic strings with $q_a = 1$, for different ALP masses, obtained with the SPA5k+B dataset. Bounds are 68\% credible intervals for detected parameters and 95\% upper limits otherwise.}
    \label{tab: Posteriors for q1 SPA5k+BAO}
\end{table}

\begin{table}[h]
    \renewcommand{\arraystretch}{1.5}
    \centering
    {\small \textbf{\boldmath $q_a = 1$, \; SPA5k+B+L}\par}
    \vspace{3pt}
    \resizebox{\textwidth}{!}{%
    \begin{tabular}{ l c c c c c c }
        \hline
        \hline
            & {\boldmath $10^{-12} \, \mathrm{eV}$}
            & {\boldmath $10^{-14} \, \mathrm{eV}$}
            & {\boldmath $10^{-16} \, \mathrm{eV}$}
            & {\boldmath $10^{-18} \, \mathrm{eV}$}
            & {\boldmath $10^{-20} \, \mathrm{eV}$}
            & {\boldmath $10^{-21} \, \mathrm{eV}$} \\
        \hline
            {\boldmath$\Omega_\mathrm{b}h^2$} & 
            $0.022534 \pm 0.000093$ & 
            $0.022528 \pm 0.000095$ & 
            $0.022501 \pm 0.000093$ & 
            $0.022498 \pm 0.000092$ & 
            $0.022498 \pm 0.000091$ & 
            $0.022501 \pm 0.000093$ \\

            {\boldmath$\Omega_\mathrm{c}h^2$} & 
            $< 0.112$ & 
            $0.066^{+0.048}_{-0.024}$ & 
            $0.101^{+0.017}_{-0.0036}$ & 
            $0.1152^{+0.0027}_{-0.00095}$ & 
            $0.11712^{+0.00086}_{-0.00070}$ & 
            $0.11743 \pm 0.00070$ \\

            {\boldmath$100\theta_\mathrm{s}$} & 
            $1.04181 \pm 0.00022$ & 
            $1.04181 \pm 0.00023$ & 
            $1.04180 \pm 0.00022$ & 
            $1.04178 \pm 0.00022$ & 
            $1.04178 \pm 0.00022$ & 
            $1.04178 \pm 0.00022$ \\

            {\boldmath$\ln(10^{10} A_\mathrm{s})$} & 
            $3.066 \pm 0.012$ & 
            $3.066 \pm 0.012$ & 
            $3.066^{+0.011}_{-0.013}$ & 
            $3.067^{+0.011}_{-0.013}$ & 
            $3.067 \pm 0.012$ & 
            $3.068^{+0.012}_{-0.013}$ \\

            {\boldmath$n_\mathrm{s}$} & 
            $0.9740 \pm 0.0030$ & 
            $0.9738 \pm 0.0029$ & 
            $0.9734 \pm 0.0030$ & 
            $0.9733 \pm 0.0030$ & 
            $0.9733 \pm 0.0029$ & 
            $0.9733 \pm 0.0031$ \\

            {\boldmath$\tau_\mathrm{reio}$} & 
            $0.0670 \pm 0.0061$ & 
            $0.0673^{+0.0058}_{-0.0066}$ & 
            $0.0675^{+0.0057}_{-0.0066}$ & 
            $0.0676^{+0.0059}_{-0.0066}$ & 
            $0.0678^{+0.0060}_{-0.0067}$ & 
            $0.0680^{+0.0061}_{-0.0069}$ \\[3pt]
        \hdashline
            {\boldmath$\Omega_{a}h^2$} & 
            $< 0.109$ & 
            $< 0.108$ & 
            $< 0.0498$ & 
            $< 0.00732$ & 
            $< 0.00175$ & 
            $< 0.00103$ \\
        \hline
        \hline
    \end{tabular}
    }
    \caption{Constraints on the cosmological parameters and on $\Omega_a h^2$ for ALPs from cosmic strings with $q_a = 1$, for different ALP masses, obtained with the SPA5k+B+L dataset. Bounds are 68\% credible intervals for detected parameters and 95\% upper limits otherwise.}
    \label{tab: Posteriors for q1 SPA5k+BAO+Plens}
\end{table}

\begin{table}[h]
    \renewcommand{\arraystretch}{1.5}
    \centering
    {\small \textbf{\boldmath $q_a = 2$, \; SPA5k+B+L}\par}
    \vspace{3pt}
    \resizebox{\textwidth}{!}{%
    \begin{tabular}{ l c c c c c c }
        \hline
        \hline
            & {\boldmath $10^{-12} \, \mathrm{eV}$}
            & {\boldmath $10^{-14} \, \mathrm{eV}$}
            & {\boldmath $10^{-16} \, \mathrm{eV}$}
            & {\boldmath $10^{-18} \, \mathrm{eV}$}
            & {\boldmath $10^{-20} \, \mathrm{eV}$}
            & {\boldmath $10^{-21} \, \mathrm{eV}$} \\
        \hline
            {\boldmath$\Omega_\mathrm{b}h^2$} & 
            $0.022530 \pm 0.000094$ & 
            $0.022528 \pm 0.000095$ & 
            $0.022521 \pm 0.000095$ & 
            $0.022506 \pm 0.000094$ & 
            $0.022500 \pm 0.000093$ & 
            $0.022501 \pm 0.000093$ \\

            {\boldmath$\Omega_\mathrm{c}h^2$} & 
            $< 0.112$ & 
            $< 0.111$ & 
            $< 0.114$ & 
            $0.107^{+0.011}_{-0.0027}$ & 
            $0.1167^{+0.0013}_{-0.00072}$ & 
            $0.11743 \pm 0.00071$ \\

            {\boldmath$100\theta_\mathrm{s}$} & 
            $1.04180 \pm 0.00022$ & 
            $1.04181 \pm 0.00023$ & 
            $1.04181 \pm 0.00022$ & 
            $1.04180^{+0.00021}_{-0.00024}$ & 
            $1.04178 \pm 0.00022$ & 
            $1.04178 \pm 0.00022$ \\

            {\boldmath$\ln(10^{10} A_\mathrm{s})$} & 
            $3.066 \pm 0.012$ & 
            $3.066^{+0.011}_{-0.013}$ & 
            $3.067 \pm 0.012$ & 
            $3.067^{+0.011}_{-0.013}$ & 
            $3.067^{+0.011}_{-0.013}$ & 
            $3.068 \pm 0.013$ \\

            {\boldmath$n_\mathrm{s}$} & 
            $0.9739 \pm 0.0030$ & 
            $0.9740 \pm 0.0029$ & 
            $0.9737 \pm 0.0029$ & 
            $0.9734 \pm 0.0030$ & 
            $0.9732 \pm 0.0030$ & 
            $0.9734 \pm 0.0031$ \\

            {\boldmath$\tau_\mathrm{reio}$} & 
            $0.0670 \pm 0.0063$ & 
            $0.0671^{+0.0058}_{-0.0065}$ & 
            $0.0674^{+0.0057}_{-0.0066}$ & 
            $0.0677^{+0.0057}_{-0.0066}$ & 
            $0.0675^{+0.0058}_{-0.0067}$ & 
            $0.0679^{+0.0060}_{-0.0068}$ \\[3pt]
        \hdashline
            {\boldmath$\Omega_{a}h^2$} & 
            $< 0.111$ & 
            $< 0.110$ & 
            $< 0.105$ & 
            $< 0.0323$ & 
            $< 0.00322$ & 
            $< 0.00109$ \\
        \hline
        \hline
    \end{tabular}
    }
    \caption{Constraints on the cosmological parameters and on $\Omega_a h^2$ for ALPs from cosmic strings with $q_a = 2$, for different ALP masses, obtained with the SPA5k+B+L dataset. Bounds are 68\% credible intervals for detected parameters and 95\% upper limits otherwise.}
    \label{tab: Posteriors for q2 SPA5k+BAO+Plens}
\end{table}

\begin{table}[h]
    \renewcommand{\arraystretch}{1.5}
    \centering
    {\small \textbf{\boldmath $q_a = 1$, \; SPA3k+B}\par}
    \vspace{3pt}
    \resizebox{\textwidth}{!}{%
    \begin{tabular}{ l c c c c c c }
        \hline
        \hline
            & {\boldmath $10^{-12} \, \mathrm{eV}$}
            & {\boldmath $10^{-14} \, \mathrm{eV}$}
            & {\boldmath $10^{-16} \, \mathrm{eV}$}
            & {\boldmath $10^{-18} \, \mathrm{eV}$}
            & {\boldmath $10^{-20} \, \mathrm{eV}$}
            & {\boldmath $10^{-21} \, \mathrm{eV}$} \\
        \hline
            {\boldmath$\Omega_\mathrm{b}h^2$} & 
            $0.022531 \pm 0.000096$ & 
            $0.022524 \pm 0.000096$ & 
            $0.022499 \pm 0.000096$ & 
            $0.022488 \pm 0.000096$ & 
            $0.022491 \pm 0.000095$ & 
            $0.022490 \pm 0.000096$ \\

            {\boldmath$\Omega_\mathrm{c}h^2$} & 
            $< 0.111$ & 
            $< 0.111$ & 
            $0.0863^{+0.031}_{-0.0089}$ & 
            $0.1126^{+0.0053}_{-0.0014}$ & 
            $0.1168^{+0.0011}_{-0.00077}$ & 
            $0.11724^{+0.00079}_{-0.00069}$ \\

            {\boldmath$100\theta_\mathrm{s}$} & 
            $1.04182 \pm 0.00022$ & 
            $1.04182 \pm 0.00022$ & 
            $1.04181 \pm 0.00023$ & 
            $1.04179 \pm 0.00023$ & 
            $1.04179 \pm 0.00022$ & 
            $1.04178 \pm 0.00023$ \\

            {\boldmath$\ln(10^{10} A_\mathrm{s})$} & 
            $3.064 \pm 0.013$ & 
            $3.063^{+0.012}_{-0.015}$ & 
            $3.064^{+0.012}_{-0.014}$ & 
            $3.064^{+0.012}_{-0.015}$ & 
            $3.064^{+0.013}_{-0.015}$ & 
            $3.065^{+0.013}_{-0.015}$ \\

            {\boldmath$n_\mathrm{s}$} & 
            $0.9738 \pm 0.0030$ & 
            $0.9738 \pm 0.0030$ & 
            $0.9733 \pm 0.0030$ & 
            $0.9731 \pm 0.0030$ & 
            $0.9731 \pm 0.0030$ & 
            $0.9733 \pm 0.0030$ \\

            {\boldmath$\tau_\mathrm{reio}$} & 
            $0.0660 \pm 0.0064$ & 
            $0.0658^{+0.0059}_{-0.0073}$ & 
            $0.0663^{+0.0059}_{-0.0070}$ & 
            $0.0663^{+0.0061}_{-0.0073}$ & 
            $0.0663^{+0.0061}_{-0.0073}$ & 
            $0.0667^{+0.0062}_{-0.0074}$ \\[3pt]
        \hdashline
            {\boldmath$\Omega_{a}h^2$} & 
            $< 0.110$ & 
            $< 0.111$ & 
            $< 0.0856$ & 
            $< 0.0149$ & 
            $< 0.00268$ & 
            $< 0.00145$ \\
        \hline
        \hline
    \end{tabular}
    }
    \caption{Constraints on the cosmological parameters and on $\Omega_a h^2$ for ALPs from cosmic strings with $q_a = 1$, for different ALP masses, obtained with the SPA3k+B dataset. Bounds are 68\% credible intervals for detected parameters and 95\% upper limits otherwise.}
    \label{tab: Posteriors for q1 SPA3k+BAO}
\end{table}

\begin{table}[h]
    \renewcommand{\arraystretch}{1.5}
    \centering
    {\small \textbf{\boldmath $q_a = 1$, \; SPA3k+B+L}\par}
    \vspace{3pt}
    \resizebox{\textwidth}{!}{%
    \begin{tabular}{ l c c c c c c }
        \hline
        \hline
            & {\boldmath $10^{-12} \, \mathrm{eV}$}
            & {\boldmath $10^{-14} \, \mathrm{eV}$}
            & {\boldmath $10^{-16} \, \mathrm{eV}$}
            & {\boldmath $10^{-18} \, \mathrm{eV}$}
            & {\boldmath $10^{-20} \, \mathrm{eV}$}
            & {\boldmath $10^{-21} \, \mathrm{eV}$} \\
        \hline
            {\boldmath$\Omega_\mathrm{b}h^2$} & 
            $0.022526 \pm 0.000096$ & 
            $0.022523 \pm 0.000096$ & 
            $0.022501 \pm 0.000094$ & 
            $0.022486 \pm 0.000096$ & 
            $0.022488 \pm 0.000094$ & 
            $0.022489 \pm 0.000097$ \\

            {\boldmath$\Omega_\mathrm{c}h^2$} & 
            $< 0.112$ & 
            $0.062^{+0.050}_{-0.040}$ & 
            $0.0911^{+0.027}_{-0.0065}$ & 
            $0.1138^{+0.0040}_{-0.0012}$ & 
            $0.11699^{+0.00096}_{-0.00072}$ & 
            $0.11738 \pm 0.00070$ \\

            {\boldmath$100\theta_\mathrm{s}$} & 
            $1.04182 \pm 0.00023$ & 
            $1.04182 \pm 0.00022$ & 
            $1.04182 \pm 0.00022$ & 
            $1.04179 \pm 0.00022$ & 
            $1.04178 \pm 0.00023$ & 
            $1.04179 \pm 0.00023$ \\

            {\boldmath$\ln(10^{10} A_\mathrm{s})$} & 
            $3.065^{+0.011}_{-0.013}$ & 
            $3.064 \pm 0.012$ & 
            $3.066^{+0.011}_{-0.013}$ & 
            $3.066 \pm 0.012$ & 
            $3.066 \pm 0.012$ & 
            $3.067^{+0.011}_{-0.013}$ \\

            {\boldmath$n_\mathrm{s}$} & 
            $0.9737 \pm 0.0030$ & 
            $0.9736 \pm 0.0030$ & 
            $0.9733 \pm 0.0030$ & 
            $0.9731 \pm 0.0030$ & 
            $0.9731 \pm 0.0031$ & 
            $0.9732 \pm 0.0031$ \\

            {\boldmath$\tau_\mathrm{reio}$} & 
            $0.0664^{+0.0058}_{-0.0066}$ & 
            $0.0662 \pm 0.0061$ & 
            $0.0672^{+0.0057}_{-0.0067}$ & 
            $0.0674^{+0.0059}_{-0.0067}$ & 
            $0.0671 \pm 0.0064$ & 
            $0.0674^{+0.0058}_{-0.0067}$ \\[3pt]
        \hdashline
            {\boldmath$\Omega_{a}h^2$} & 
            $< 0.110$ & 
            $< 0.109$ & 
            $< 0.0764$ & 
            $< 0.0111$ & 
            $< 0.00213$ & 
            $< 0.00122$ \\
        \hline
        \hline
    \end{tabular}
    }
    \caption{Constraints on the cosmological parameters and on $\Omega_a h^2$ for ALPs from cosmic strings with $q_a = 1$, for different ALP masses, obtained with the SPA3k+B+L dataset. Bounds are 68\% credible intervals for detected parameters and 95\% upper limits otherwise.}
    \label{tab: Posteriors for q1 SPA3k+BAO+Plens}
\end{table}

\begin{table}[h]
    \renewcommand{\arraystretch}{1.5}
    \centering
    {\small \textbf{\boldmath $q_a = 2$, \; SPA3k+B+L}\par}
    \vspace{3pt}
    \resizebox{\textwidth}{!}{%
    \begin{tabular}{ l c c c c c c }
        \hline
        \hline
            & {\boldmath $10^{-12} \, \mathrm{eV}$}
            & {\boldmath $10^{-14} \, \mathrm{eV}$}
            & {\boldmath $10^{-16} \, \mathrm{eV}$}
            & {\boldmath $10^{-18} \, \mathrm{eV}$}
            & {\boldmath $10^{-20} \, \mathrm{eV}$}
            & {\boldmath $10^{-21} \, \mathrm{eV}$} \\
        \hline
            {\boldmath$\Omega_\mathrm{b}h^2$} & 
            $0.022527 \pm 0.000098$ & 
            $0.022524 \pm 0.000097$ & 
            $0.022520 \pm 0.000098$ & 
            $0.022499 \pm 0.000097$ & 
            $0.022496 \pm 0.000092$ & 
            $0.022492 \pm 0.000098$ \\

            {\boldmath$\Omega_\mathrm{c}h^2$} & 
            $< 0.111$ & 
            $< 0.112$ & 
            $0.065^{+0.049}_{-0.023}$ & 
            $0.0989^{+0.019}_{-0.0040}$ & 
            $0.1163^{+0.0016}_{-0.00077}$ & 
            $0.11739 \pm 0.00068$ \\

            {\boldmath$100\theta_\mathrm{s}$} & 
            $1.04181 \pm 0.00022$ & 
            $1.04181 \pm 0.00023$ & 
            $1.04181 \pm 0.00022$ & 
            $1.04180 \pm 0.00023$ & 
            $1.04179 \pm 0.00022$ & 
            $1.04179 \pm 0.00022$ \\

            {\boldmath$\ln(10^{10} A_\mathrm{s})$} & 
            $3.065^{+0.011}_{-0.013}$ & 
            $3.065^{+0.011}_{-0.013}$ & 
            $3.065^{+0.011}_{-0.013}$ & 
            $3.066^{+0.011}_{-0.013}$ & 
            $3.066^{+0.011}_{-0.013}$ & 
            $3.066^{+0.012}_{-0.014}$ \\

            {\boldmath$n_\mathrm{s}$} & 
            $0.9739 \pm 0.0030$ & 
            $0.9737 \pm 0.0030$ & 
            $0.9736 \pm 0.0030$ & 
            $0.9730 \pm 0.0030$ & 
            $0.9730 \pm 0.0030$ & 
            $0.9731 \pm 0.0031$ \\

            {\boldmath$\tau_\mathrm{reio}$} & 
            $0.0663^{+0.0057}_{-0.0066}$ & 
            $0.0664^{+0.0057}_{-0.0068}$ & 
            $0.0667^{+0.0057}_{-0.0068}$ & 
            $0.0672^{+0.0058}_{-0.0068}$ & 
            $0.0671^{+0.0057}_{-0.0067}$ & 
            $0.0672^{+0.0059}_{-0.0070}$ \\[3pt]
        \hdashline
            {\boldmath$\Omega_{a}h^2$} & 
            $< 0.109$ & 
            $< 0.109$ & 
            $< 0.109$ & 
            $< 0.0546$ & 
            $< 0.00416$ & 
            $< 0.00119$ \\
        \hline
        \hline
    \end{tabular}
    }
    \caption{Constraints on the cosmological parameters and on $\Omega_a h^2$ for ALPs from cosmic strings with $q_a = 2$, for different ALP masses, obtained with the SPA3k+B+L dataset. Bounds are 68\% credible intervals for detected parameters and 95\% upper limits otherwise.}
    \label{tab: Posteriors for q2 SPA3k+BAO+Plens}
\end{table}
\bibliographystyle{JHEP}
\bibliography{biblio}

\end{document}